\documentclass[%
 reprint,
superscriptaddress,
nofootinbib,
 amsmath,amssymb,
 aps,
 prd,
 showkeys,
floatfix,
]{revtex4-2}

\usepackage{booktabs}
\usepackage{mathtools}
\usepackage{graphicx}
\usepackage{dcolumn}
\usepackage{bm}
\usepackage{amssymb}   
\usepackage{aas_macros}
\usepackage{hyperref}
\usepackage{accents}
\usepackage{multirow}
\usepackage[caption=false]{subfig}

\newcommand{\HI}{H\,{\sc i}}
\newcommand*\conj[1]{\overline{#1}}

\newcommand\fs{\hbox{$.\!\!^{\mathrm s}$}}
\newcommand\fdg{\hbox{$.\!\!^\circ$}}

\newcommand\farcs{\hbox{$.\!\!^{\prime\prime}$}}
\newcommand\arcmin{\hbox{$^\prime$}}

\newcommand\arcdeg{\mbox{$^\circ$}}

\makeatletter
\renewcommand\paragraph{\@startsection{paragraph}{4}{\z@}%
            {-2.5ex\@plus -1ex \@minus -.25ex}%
            {1.25ex \@plus .25ex}%
            {\normalfont\normalsize\itshape}}
\makeatother
\begin{document}

\preprint{APS/123-QED}

\title{Detection of Cosmic Structures using the Bispectrum Phase. I. \\ Mathematical Foundations}
\author{Nithyanandan Thyagarajan}
\thanks{Nithyanandan Thyagarajan is a Jansky Fellow of the National Radio Astronomy Observatory}
\email{\\ t\_nithyanandan@nrao.edu, nithyanandan.t@gmail.com}
\homepage{\\ https://tnithyanandan.wordpress.com/}
\affiliation{National Radio Astronomy Observatory, Socorro, NM 87801, USA}
\affiliation{Arizona State University, School of Earth and Space Exploration, Tempe, AZ 85287, USA}

\author{Chris L. Carilli}%
\affiliation{National Radio Astronomy Observatory, Socorro, NM 87801, USA}
\affiliation{Astrophysics Group, Cavendish Laboratory, University of Cambridge, Cambridge CB3 0HE, UK}

\date{\today}

\begin{abstract}

Many low-frequency radio interferometers are aiming to detect very faint spectral signatures from structures at cosmological redshifts, particularly of neutral Hydrogen using its characteristic 21~cm spectral line. Due to the very high dynamic range needed to isolate these faint spectral fluctuations from the very bright foregrounds, spectral systematics from the instrument or the analysis, rather than thermal noise, are currently limiting their sensitivity. Failure to achieve a spectral calibration of the instrument with fractional inaccuracy $\lesssim 10^{-5}$ will make the detection of the critical cosmic signal unlikely. The bispectrum phase from interferometric measurements is largely immune to this calibration issue. We present a basis to explore the nature of the bispectrum phase in the limit of small spectral fluctuations. We establish that these fluctuations measure the intrinsic dissimilarity in the transverse structure of the cosmic signal relative to the foregrounds, expressed as rotations in the underlying phase angle. Their magnitude is related to the strength of the cosmic signal relative to the foregrounds. Using a range of sky models, we detail the behavior of the bispectrum phase fluctuations using standard Fourier-domain techniques and find it comparable to existing approaches, with a few key differences. Foreground contamination from \textit{mode-mixing} between the transverse and line-of-sight dimensions is more pronounced than in existing approaches because the bispectrum phase is a product of three individual interferometric phases. The multiplicative coupling of foregrounds in the bispectrum phase fluctuations results in the mixing of foreground signatures with that of the cosmic signal. We briefly outline a variation of this approach to avoid extensive mode-mixing. Despite its limitations, the interpretation of results using the bispectrum phase is possible with forward-modeling. Importantly, it is an independent and a viable alternative to existing approaches.

\end{abstract}

\maketitle


\section{Introduction}

The formation and evolution of large-scale structure in the high-redshift Universe ($1\lesssim z\lesssim 100$) has been largely underexplored. Probing the early Universe using spectral line tracers on cosmological scales appears promising and could be very rewarding scientifically to both understand these processes and their effects on shaping the astrophysical evolution of the Universe \cite[see e.g.][]{sun72,sco90,gne97,mad97,sha99,toz00,ili02,fan02,gne04,fan06,fur06,bar07,mor10,pri12}. Examples include tomographic mapping of the neutral Hydrogen (\HI) using the redshifted 21~cm spectral line from the electron spin-flip transition during the \textit{Cosmic Dark Ages} ($z\gtrsim 30$), the \textit{Cosmic Dawn} ($15\lesssim z\lesssim 30$), the \textit{Epoch of Reionization} (EoR; $6\lesssim z\lesssim 15$), and the periods when \textit{Dark Energy} started becoming significant ($1\lesssim z\lesssim 3$) and eventually dominant ($z<1$).

Rapid advances in low radio frequency instrumentation has made it possible for a number of experiments including the the Murchison Widefield Array \citep[MWA;][]{lon09,bow13,tin13,bea19}, the Donald~C.~Backer Precision Array for Probing the Epoch of Reionization \citep[PAPER;][]{par10}, the Low Frequency Array \citep[LOFAR;][]{van13}, the Giant Metrewave Radio Telescope EoR experiment \citep[GMRT;][]{pac13}, the Hydrogen Epoch of Reionization Array \citep[HERA;][]{deb17}, the Square Kilometre Array \citep[SKA;][]{mel13}, the Canadian Hydrogen Intensity Mapping Experiment \citep[CHIME;][]{ban14}, and the Hydrogen Intensity and Real-time Analysis eXperiment \citep[HIRAX;][]{new16} to attempt detecting the cosmic \HI\ structures in these epochs using the redshifted 21~cm spectral line with sufficient sensitivity \cite[see e.g.][]{bea13,thy13}. 

With the requirement to isolate very weak spectral signatures in the presence of very bright foreground objects, these experiments are faced with a tremendous challenge of requiring extreme-fidelity spectral calibration \cite{dat10,barry16,tro16,patil17} with fractional inaccuracy typically under $\lesssim 10^{-5}$, failing which the miscalibration will leak sufficient power from the bright foregrounds and contaminate the spectral signatures of the cosmic signal, thereby rendering this critical detection impossible. While advanced calibration methods are being investigated to address the calibration challenge \citep[see e.g.][]{ewa17,sie17,dil18,oro19}, a new and independent approach to detecting the spectral signatures from cosmic structures using the bispectrum phase was presented recently \cite{thy18}, which has the distinct advantage that it is largely impervious to antenna-based calibration and the errors therein. This property has been investigated in detail \cite{jen58,kul89,tay99,tho01,mon07b}. The bispectrum phase intrinsically measures the symmetry about a point and is invariant to translation \cite{mon07b}. Indeed, the bispectrum phase has been exploited successfully in interferometric imaging experiments where calibration is extremely challenging, such as in deciphering complex structures on stellar surfaces and their surroundings \cite[][and references therein]{mon06,mon07a}, and in the Event Horizon Telescope (EHT) imaging of the shadow of the supermassive black hole at the center of M87 \cite{eht19-1,eht19-2,eht19-3,eht19-4,eht19-5,eht19-6}. 

Most of the aforementioned applications of bispectrum phase were used for imaging which is restricted to the transverse plane of the sky. Following up on the idea presented in \cite{thy18}, we explore and exploit new  properties of the bispectrum phase while applying it to the spectral (or the frequency) axis, which in cosmological spectral line observations is typically the line-of-sight dimension of the sky. This paper is one in a series of related papers, the others being \cite{thy18,car18,car20,thy20b}, and lays the mathematical foundations for our understanding and application of the bispectrum phase approach for spectral line observations where information about the desired signal can be extracted through its distinct spectral signatures. In a companion paper \cite[][hereafter Paper II]{thy20b}, we present the first results from applying this technique to a small sample of data obtained with the HERA telescope. 

This paper is organized as follows. In \S\ref{sec:complex-algebra}, we present a basis using simple complex algebra to simplify the mathematical understanding of phase fluctuations in the limit of small perturbations. We apply this simplification first to phases in the interferometric two-point correlations (or visibilities) in \S\ref{sec:visphase-math}. We then extend this formalism to the bispectrum phase in \S\ref{sec:bsp-math}. In \S\ref{sec:examples}, we use a range of purely hypothetical to realistic examples to identify and illustrate the relationship and correspondence, and the benefits and limitations, of using our bispectrum phase approach relative to  other existing approaches that use visibilities in distinguishing the spectral signatures from the cosmic signal. In \S\ref{sec:mode-mixing}, we demonstrate the presence and effects of \textit{mode-mixing} (the coupling of spatial modes along the line-of-sight direction with those in the transverse plane) in our approach similar to that in existing approaches. In \S\ref{sec:bsp-angle-approach}, we briefly outline an alternate approach that could potentially mitigate contamination from \textit{mode-mixing} effects. In \S\ref{sec:fg-impact}, we discuss the impact of the intrinsic spectral characteristics of the foregrounds on the spectral signatures of the cosmic spectral line signal and contrast it with existing approaches. We provide a summary of the mathematical formalism of this approach along with its benefits and limitations in \S\ref{sec:summary}. While we use examples relating to the detection of \HI\ in the intergalactic medium during the EoR, the formalism and the conclusions presented here are generically applicable to other experiments and science cases as well.  

\section{Linear Order Fluctuations on Complex Vectors}\label{sec:complex-algebra}

Consider complex numbers $Z_j$, with amplitudes $|Z_j|$ and arguments $\theta_j$ in the complex plane, $\mathbb{C}$, such that $Z_j = |Z_j|\,e^{i\theta_j}$, for $j=0, 1, 2, \ldots n$. The real and imaginary parts are denoted by $\Re\{\cdot\}$ and $\Im\{\cdot\}$ respectively. $\conj{Z}$ denotes the complex conjugate of $Z$. 

In this paper, we denote $Z_0$ as the reference complex vector, and $Z_j$ for $1\le j\le n$ as perturbations over $Z_0$. Let $Z_\Sigma = \sum_{j=0}^n\,Z_j$ denote the resultant complex vector. Throughout this paper, we will often invoke that $|Z_j|/|Z_0| \ll 1$ for $j\ge 1$, so that only linear-order terms in $|Z_j|/|Z_0|$ will be retained while neglecting higher-order terms. 

\subsection{Small Perturbations from a Single Cause}

For a small perturbation arising from a single cause ($n=1$), say only one of thermal noise or spectral line fluctuations, $Z_\Sigma = Z_0 + Z_1$. The amplitude and argument of the resultant are:
\begin{align}
    |Z_\Sigma|^2 &= |Z_0|^2 + |Z_1|^2 + 2 \Re\{Z_0\,\conj{Z_1}\}, \\
    \theta_\Sigma &= \theta_0 + \delta\theta_0,
\end{align}
where,
\begin{equation}
    \tan\delta\theta_0 = \frac{|Z_1|\sin(\theta_1-\theta_0)}{|Z_0| + |Z_1|\cos(\theta_1-\theta_0)}.
\end{equation}

Assuming the perturbation is small, $|Z_1| \ll |Z_0|$, using Taylor-series expansion and small-angle approximation, and retaining only up to linear-order terms in $|Z_1|/|Z_0|$,
\begin{align}
    |Z_\Sigma| &\approx |Z_0|\,\left(1+\frac{|Z_1|}{|Z_0|}\cos(\theta_1-\theta_0)\right) \label{eqn:amp-lin-1-a} \\
    &= |Z_0|\,\left(1+\Re\left\{\frac{Z_1}{Z_0}\right\}\right), \label{eqn:amp-lin-1-b}
\end{align}
and,
\begin{align}
    \tan\delta\theta_0 \approx \delta\theta_0 &\approx \frac{|Z_1|}{|Z_0|}\sin(\theta_1-\theta_0) = \Im\left\{\frac{Z_1}{Z_0}\right\}. \label{eqn:ang-lin-1}
\end{align}
Equation~(\ref{eqn:ang-lin-1}) establishes that the fluctuation in the angle of the reference complex vector depends on $|Z_1|/|Z_0|$ and on the relative angle between the two, $\theta_1-\theta_0$. Hence, if the fluctuation has zero magnitude ($|Z_1|=0$) or if its angle is exactly aligned with the reference vector ($\theta_1=\theta_0$), it will cause no perturbation in angle. Thus, perturbations in the underlying phase angle are an intrinsic measure of the dissimilarity of the perturbing signal relative to the underlying signal, and the magnitude of the phase angle perturbation depends on the ratio between the two.

\subsection{Small Perturbations from Multiple Causes}\label{sec:multiple-linear-perturbations}

The relationships can be generalized and extended in case of multiple causes of perturbations such as simultaneous presence of thermal noise and spectral line fluctuations. In the presence of multiple sources of perturbations, the same small-angle and linear-order approximation yields:
\begin{align}
    |Z_\Sigma| &\approx \sum_{j=0}^n\, |Z_j|\cos(\theta_j-\theta_0) \\
    &= |Z_0|\sum_{j=0}^n\,\frac{|Z_j|}{|Z_0|}\cos(\theta_j-\theta_0) \\
    &= |Z_0|\sum_{j=0}^n\,\Re\left\{\frac{Z_j}{Z_0}\right\} = |Z_0|\,\Re\left\{\sum_{j=0}^n\,\frac{Z_j}{Z_0}\right\},
\end{align}
and,
\begin{align}
    \delta\theta_0 &\approx \sum_{j=0}^n\,\frac{|Z_j|}{|Z_0|}\sin(\theta_j-\theta_0) \\ 
    &= \sum_{j=0}^n\,\Im\left\{\frac{Z_j}{Z_0}\right\} = \Im\left\{\sum_{j=0}^n\,\frac{Z_j}{Z_0}\right\}
\end{align}

It may be noted that when these perturbations arise from Gaussian noise and are small, wherein $|Z_j|/|Z_0| \ll 1$, the distribution of $\delta\theta_0$ can also be well approximated by a Gaussian distribution \cite{cra89}.  

\section{Spectral Line Fluctuations in Visibility Phase}\label{sec:visphase-math}

In this section, we develop the mathematical formalism on the interferometric visibility phases that will be later extended to the bispectrum phase. Let $V_p(f)$ denote the spectrum of visibilities of a triad of a baseline (antenna spacing) vectors, $\boldsymbol{b}_p$, where, $p=\{1,2,3\}$ indexes the baselines comprising the triad. Then, a measured visibility (after calibration) can be written as:
\begin{align}
    V_p^\textrm{m}(f) &= V_p^\textrm{T}(f) + V_p^\textrm{N}(f),
\end{align}
where, superscripts $\textrm{m}$, $\textrm{T}$, and $\textrm{N}$ denote the \textit{measured}, \textit{true sky}, and \textit{noise} components respectively. The \textit{true sky} visibility can be further decomposed as:
\begin{align}
    V_p^\textrm{T}(f) &= V_p^\textrm{F}(f) + V_p^\textrm{L}(f),
\end{align}
where, $\textrm{F}$ denotes the foregrounds and $\textrm{L}$ denotes the cosmological spectral line signal of interest. Hence,
\begin{align}
    V_p^\textrm{m}(f) &= V_p^\textrm{F}(f) + V_p^\textrm{L}(f) + V_p^\textrm{N}(f).
\end{align}

In these visibilities, let $\phi_p^\textrm{m}(f)$ be the measured interferometric phase angle, $\phi_p^\textrm{T}(f)$ be the uncorrupted interferometric phase angle relating to the true sky, $\phi_p^\textrm{F}(f)$ be the uncorrupted interferometric phase angle from foregrounds, $\delta\phi_p^\textrm{L}(f)$ be the perturbation caused by cosmic structures of interest probed by the spectral line to the uncorrupted foreground interferometric phase angle, and $\delta\phi_p^\textrm{N}(f)$ be the perturbation to the uncorrupted interferometric phase angle caused by thermal noise. If the perturbations to visibility are small, $|V_p^\textrm{L}(f)|\ll |V_p^\textrm{F}(f)|$ and $|V_p^\textrm{N}(f)|\ll |V_p^\textrm{F}(f)|$, then the approximations in \S\ref{sec:multiple-linear-perturbations} can be employed. Identifying $V_p^\textrm{F}$, $V_p^\textrm{L}$, and $V_p^\textrm{N}$ to be respectively $Z_0$, $Z_1$, and $Z_2$ in \S\ref{sec:multiple-linear-perturbations}, the perturbations to the visibility phase angle from the cosmic structures and noise are, respectively:
\begin{align}
    \delta\phi_p^\textrm{L}(f) &\approx \Im\left\{\frac{V_p^\textrm{L}(f)}{V_p^\textrm{F}(f)}\right\}, \label{eqn:visphase-line-perturbations} \\
    \textrm{and,}\quad\delta\phi_p^\textrm{N}(f) &\approx \Im\left\{\frac{V_p^\textrm{N}(f)}{V_p^\textrm{F}(f)}\right\}. \label{eqn:visphase-noise-perturbations}
\end{align}
Thus fluctuations in the visibility phase angles are roughly linearly proportional to the fluctuations in the visibilities themselves and inversely proportional to the foreground component of the visibilities under a first-order approximation. Then,
\begin{align}
    e^{i\phi_p^\textrm{m}(f)} &= e^{i(\phi_p^\textrm{F}(f) + \delta\phi_p^\textrm{L}(f) + \delta\phi_p^\textrm{N}(f))} \\
    &\approx e^{i\phi_p^\textrm{F}(f)}\left[1+i\left(\delta\phi_p^\textrm{L}(f)+\delta\phi_p^\textrm{N}(f)\right)\right] \\
    &\approx e^{i\phi_p^\textrm{F}(f)} \Biggl[1+\frac{1}{2}\left(\frac{V_p^\textrm{L}(f)}{V_p^\textrm{F}(f)}-\frac{\conj{V_p^\textrm{L}}(f)}{\conj{V_p^\textrm{F}}(f)}\right) \nonumber\\ 
    &\qquad\qquad\qquad +\frac{1}{2}\left(\frac{V_p^\textrm{N}(f)}{V_p^\textrm{F}(f)}-\frac{\conj{V_p^\textrm{N}}(f)}{\conj{V_p^\textrm{F}}(f)}\right)\Biggr], \label{eqn:linearized-visphase-visratio}
\end{align}
where, we have used $|\delta\phi_p^\textrm{L}(f)| \ll 1$ and $|\delta\phi_p^\textrm{N}(f)| \ll 1$, and $\Im\{Z\} = (Z-\conj{Z})/2i$, in Eqs.~(\ref{eqn:visphase-line-perturbations}) and (\ref{eqn:visphase-noise-perturbations}). Thus the visibility phase angle fluctuations $\delta\phi_p^\textrm{L}(f)$ and $\delta\phi_p^\textrm{N}(f)$ are indicators of the spectral line strength to the foreground continuum ratio and the noise to foreground continuum ratio respectively. 

If a model for $|V_p^\textrm{F}(f)|$ is available, it can be used to extract partial information about $V_p^\textrm{L}(f)$. Let $|\widehat{V}_p^\textrm{F}(f)|$ denote some empirical model of the true sky-based foreground visibility amplitude, $|V_p^\textrm{F}(f)|$. Then, we can reconstruct an estimate of the measured (calibrated) visibility as:
\begin{align}
    \widehat{V}_p^\textrm{m}(f) &= |\widehat{V}_p^\textrm{F}(f)| \, e^{i\phi_p^\textrm{m}(f)} \nonumber\\
    &= |\widehat{V}_p^\textrm{F}(f)| \, e^{i\phi_p^\textrm{F}(f)} \Biggl[1+\frac{1}{2}\left(\frac{V_p^\textrm{L}(f)}{V_p^\textrm{F}(f)}-\frac{\conj{V_p^\textrm{L}}(f)}{\conj{V_p^\textrm{F}}(f)}\right) \nonumber\\ 
    &\qquad\qquad\qquad\quad +\frac{1}{2}\left(\frac{V_p^\textrm{N}(f)}{V_p^\textrm{F}(f)}-\frac{\conj{V_p^\textrm{N}}(f)}{\conj{V_p^\textrm{F}}(f)}\right)\Biggr] \nonumber\\
    &= \widehat{V}_p^\textrm{F}(f)\Biggl[1+\frac{1}{2}\left(\frac{V_p^\textrm{L}(f)}{V_p^\textrm{F}(f)}-\frac{\conj{V_p^\textrm{L}}(f)}{\conj{V_p^\textrm{F}}(f)}\right) \nonumber\\ 
    &\qquad\qquad\qquad\quad +\frac{1}{2}\left(\frac{V_p^\textrm{N}(f)}{V_p^\textrm{F}(f)}-\frac{\conj{V_p^\textrm{N}}(f)}{\conj{V_p^\textrm{F}}(f)}\right)\Biggr] \nonumber\\
    &= \widehat{V}_p^\textrm{F}(f) + \widehat{V}_p^\textrm{L}(f) + \widehat{V}_p^\textrm{N}(f), \label{eqn:visphase-equivalence}
\end{align}
where, 
\begin{align}
    \widehat{V}_p^\textrm{L}(f) &= \frac{\widehat{V}_p^\textrm{F}(f)}{2} \left(\frac{V_p^\textrm{L}(f)}{V_p^\textrm{F}(f)}-\frac{\conj{V_p^\textrm{L}}(f)}{\conj{V_p^\textrm{F}}(f)}\right) \\
    \textrm{and,}\quad \widehat{V}_p^\textrm{N}(f) &= \frac{\widehat{V}_p^\textrm{F}(f)}{2} \left(\frac{V_p^\textrm{N}(f)}{V_p^\textrm{F}(f)}-\frac{\conj{V_p^\textrm{N}}(f)}{\conj{V_p^\textrm{F}}(f)}\right).
\end{align}
This will not yield a perfect recovery of the fluctuating components because the phase angle fluctuations are related by the imaginary portion of the ratio of the fluctuating component to the foreground and hence only yields a partial recovery, statistically $\sim 50$\% of the spatial information content. Nevertheless, it can be useful in recovering roughly half the power in the fluctuations.

Note that the use of calibrated interferometric phase angle $\phi_p^\textrm{m}(f)$ is required in Equation~(\ref{eqn:visphase-equivalence}) and could not have been substituted by the uncalibrated interferometric phase angle because the latter will result in partial to complete loss of recovery of the visibilities if the measured interferometric phase is uncorrected for the phase corruption of the wavefront introduced by the antenna and the ionosphere. On the other hand, the use of corrected phase carries the burden of having performed extremely accurate calibration, typically with fractional inaccuracy required to be $\lesssim 10^{-5}$. However, the bispectrum phase has the interesting property that it is independent of direction-independent antenna-based calibration and errors therein. The following sections extend this treatment of interferometric phase on visibilities to examine the usefulness of the bispectrum phase while using the raw uncalibrated measurements.

\section{Spectral Line Fluctuations in the Bispectrum Phase}\label{sec:bsp-math}

The measured complex bispectrum is written as the product (over index $p$) of the measured visibilities (may or may not be calibrated):
\begin{align}
B_\nabla^\textrm{m}(f) &= |B_\nabla^\textrm{m}(f)|\,e^{i\phi_\nabla^\textrm{m}(f)} = \prod_{p=1}^3 \,V_p^\textrm{m}(f)\nonumber\\
	& = G_\nabla(f)\,\prod_{p=1}^3 \, \Bigl[V_p^\textrm{F}(f)+V_p^\textrm{L}(f)+V_p^\textrm{N}(f)\Bigr], \label{eqn:bispectrum-components}
\end{align}
where, $G_\nabla(f)$ denotes the lumped product of the various direction-independent antenna-based gains associated with the triad, and $|B_\nabla^\textrm{m}(f)|$ and $\phi_\nabla^\textrm{m}(f)$ denote the amplitude and the phase angle of the measured bispectrum, respectively. 

The measured bispectrum phase angle, which is independent of the direction-independent antenna-based gains \citep{jen58}, is:
\begin{align}
    \phi_\nabla^\textrm{m}(f) &= \sum_{p=1}^3 \, \phi_p^\textrm{m}(f) = \sum_{p=1}^3 \, \Bigl[\phi_p^\textrm{F}(f) + \delta\phi_p^\textrm{L}(f) + \delta\phi_p^\textrm{N}(f)\Bigr] \\
    &= \sum_{p=1}^3 \, \phi_p^\textrm{F}(f) + \sum_{p=1}^3 \, \delta\phi_p^\textrm{L}(f) + \sum_{p=1}^3 \, \delta\phi_p^\textrm{N}(f) \label{eqn:baseline-phase-components}\\
    &= \phi_\nabla^\textrm{F}(f) + \delta\phi_\nabla^\textrm{L}(f) + \delta\phi_\nabla^\textrm{N}(f), \label{eqn:bispectrum-phase-components}
\end{align}
where, $\phi_\nabla^\textrm{F}(f)=\sum_{p=1}^3 \, \phi_p^\textrm{F}(f)$ is the bispectrum phase angle from the foreground structures, $\delta\phi_\nabla^\textrm{L}(f)=\sum_{p=1}^3 \, \delta\phi_p^\textrm{L}(f)$ is the perturbation to the bispectrum phase angle caused by the presence of cosmic structures, and $\delta\phi_\nabla^\textrm{N}(f)=\sum_{p=1}^3 \, \delta\phi_p^\textrm{N}(f)$ is the perturbation to the bispectrum phase angle caused by the presence of thermal noise. 

We assume throughout the paper that during an observation the phase center coincides with the pointing center (boresight) of the antenna. For a transit array like HERA, this is typically the zenith. However, the discussion and examples presented here can be generalized to a tracking telescope with arbitrary pointing and phase centers as well.

The following section relates these phase fluctuations to the spatial coherence function corresponding to the cosmic structures and thermal noise. For convenience throughout, we drop the explicit dependence on $f$ unless specified. 

\subsection{Bispectrum phase in the limit of small perturbations}\label{sec:bsp-linear-perturbations}

The measured bispectrum in Equation~(\ref{eqn:bispectrum-components}) can be expanded as:
\begin{align}
    |B_\nabla^\textrm{m}|\,e^{i\phi_\nabla^\textrm{m}} &= \prod_{p=1}^3 \, V_p^\textrm{F} \nonumber\\
	&\quad + V_1^\textrm{F}\,V_2^\textrm{F}\,V_3^\textrm{L} + V_1^\textrm{F}\,V_3^\textrm{F}\,V_2^\textrm{L} + V_2^\textrm{F}\,V_3^\textrm{F}\,V_1^\textrm{L}\nonumber\\
    &\quad + V_1^\textrm{F}\,V_2^\textrm{F}\,V_3^\textrm{N} + V_1^\textrm{F}\,V_3^\textrm{F}\,V_2^\textrm{N} + V_2^\textrm{F}\,V_3^\textrm{F}\,V_1^\textrm{N} \nonumber\\    
    &\quad + \textrm{higher-order terms in}\, V^\textrm{L}\, \textrm{and}\, V^\textrm{N}.
\end{align}
Here, we have neglected the gain terms lumped into $G_\nabla(f)$ as they multiply across all the terms and are irrelevant for the phase angles and the fluctuations therein.

By keeping terms only up to linear order in $V_p^\textrm{L}$ and $V_p^\textrm{N}$, we can infer the first-order perturbation to the foreground bispectrum, $B_\nabla^\textrm{F}=\prod_{p=1}^3 \, V_p^\textrm{F}$, arising from cosmic structures and thermal noise, respectively, as:
\begin{align}
    B_\nabla^\textrm{L} &\approx V_1^\textrm{F}\,V_2^\textrm{F}\,V_3^\textrm{L} + V_1^\textrm{F}\,V_3^\textrm{F}\,V_2^\textrm{L} + V_2^\textrm{F}\,V_3^\textrm{F}\,V_1^\textrm{L}, \\
    \textrm{and,}\,\, B_\nabla^\textrm{N} &\approx V_1^\textrm{F}\,V_2^\textrm{F}\,V_3^\textrm{N} + V_1^\textrm{F}\,V_3^\textrm{F}\,V_2^\textrm{N} + V_2^\textrm{F}\,V_3^\textrm{F}\,V_1^\textrm{N}.
\end{align}
Identifying $B_\nabla^\textrm{F}$, $B_\nabla^\textrm{L}$, and $B_\nabla^\textrm{N}$ to be respectively $Z_0$, $Z_1$, and $Z_2$ in \S\ref{sec:multiple-linear-perturbations}, the perturbations to the foreground bispectrum phase from the cosmic structures and noise are, respectively:
\begin{align}
    \delta\phi_\nabla^\textrm{L} &\approx \Im\left\{\sum_{p=1}^3 \, \frac{V_p^\textrm{L}}{V_p^\textrm{F}}\right\}, \label{eqn:bispectrum-line-perturbations-1} \\
    \textrm{and,}\quad\delta\phi_\nabla^\textrm{N} &\approx \Im\left\{\sum_{p=1}^3 \, \frac{V_p^\textrm{N}}{V_p^\textrm{F}}\right\}. \label{eqn:bispectrum-noise-perturbations-1}
\end{align}
This suggests that fluctuations in the bispectrum phase are approximately linearly proportional to the fluctuations in visibilities under a first-order approximation.

Alternatively, from Eqs.~(\ref{eqn:baseline-phase-components}) and (\ref{eqn:bispectrum-phase-components}), 
\begin{align}
    \delta\phi_\nabla^\textrm{L}(f) &= \sum_{p=1}^3 \, \delta\phi_p^\textrm{L}(f) \approx \sum_{p=1}^3 \, \Im\left\{\frac{V_p^\textrm{L}}{V_p^\textrm{F}}\right\}, \label{eqn:bispectrum-line-perturbations-2} \\
    \textrm{and,}\quad \delta\phi_\nabla^\textrm{N}(f) &= \sum_{p=1}^3 \, \delta\phi_p^\textrm{N}(f) \approx \sum_{p=1}^3 \, \Im\left\{\frac{V_p^\textrm{N}}{V_p^\textrm{F}}\right\}, \label{eqn:bispectrum-noise-perturbations-2}
\end{align}
which are identical to Eqs.~(\ref{eqn:bispectrum-line-perturbations-1}) and (\ref{eqn:bispectrum-noise-perturbations-1}). Effectively, these bispectrum phase fluctuations are a reasonable instrument-independent, and a robust true-sky measure of the dissimilarity of the cosmic structures relative to the foregrounds in the transverse plane of the sky, whose magnitude depends on the ratio between the two.

\subsection{Relation between Bispectrum Phase Fluctuations and Visibilities}\label{sec:bsp-vis-relation}

In the limit of small fluctuations, $|\delta\phi_p^\textrm{L}|, |\delta\phi_p^\textrm{N}| \ll 1$, the bispectrum phase can be expressed using Taylor-series expansion to linear-order terms in $\delta\phi_\nabla^\textrm{L}$ and $\delta\phi_\nabla^\textrm{N}$ as:
\begin{align}
    e^{i\phi_\nabla^\textrm{m}} &= e^{i(\phi_\nabla^\textrm{F}+\delta\phi_\nabla^\textrm{L}+\delta\phi_\nabla^\textrm{N})} \\
    &\approx e^{i\phi_\nabla^\textrm{F}}\left[1+i(\delta\phi_\nabla^\textrm{L}+\delta\phi_\nabla^\textrm{N})\right] \label{eqn:linearized-bispectrum-phase}
\end{align}
Thus, the perturbed angles, $\delta\phi_\nabla^\textrm{L}$ and $\delta\phi_\nabla^\textrm{N}$, and also $e^{i\phi_\nabla^\textrm{m}}$ contain terms up to linear order in $V_p^\textrm{L}/V_p^\textrm{F}$ and $V_p^\textrm{N}/V_p^\textrm{F}$. From Eqs.~(\ref{eqn:bispectrum-line-perturbations-1}) and (\ref{eqn:bispectrum-noise-perturbations-1}), Equation~(\ref{eqn:linearized-bispectrum-phase}) can be expressed as:
\begin{align}
    e^{i\phi_\nabla^\textrm{m}} &\approx e^{i\phi_\nabla^\textrm{F}}\Biggl[1+\frac{1}{2}\sum_{p=1}^3\,\left(\frac{V_p^\textrm{L}}{V_p^\textrm{F}}-\frac{\conj{V_p^\textrm{L}}}{\conj{V_p^\textrm{F}}}\right) \nonumber \\ 
    &\qquad\qquad +  \frac{1}{2}\sum_{p=1}^3\,\left(\frac{V_p^\textrm{N}}{V_p^\textrm{F}}-\frac{\conj{V_p^\textrm{N}}}{\conj{V_p^\textrm{F}}}\right)\Biggr]. \label{eqn:linearized-bispectrum-phase-visratio}
\end{align}
Hence, the bispectrum phase angle fluctuations, $\delta\phi_\nabla^\textrm{L}$ and $\delta\phi_\nabla^\textrm{N}$, measure the dissimilarity relative to the foregrounds with magnitudes given by the spectral line to foreground continuum ratio, and noise-to-foreground continuum ratio, respectively. This is very similar to the nature of the visibility phase angle fluctuations. 

In principle, we could simply use $\phi_\nabla^\textrm{m}$ as the mathematical quantity of interest instead of the complex Eulerian representation $e^{i\phi_\nabla^\textrm{m}}$, that will be \textit{delay-transformed} as we have in the rest of the paper. However, in practice $\phi_\nabla^\textrm{m}$ will be noisy and may contain discontinuities at $\pm\pi$ due to phase angle wrapping. Including such discontinuities in a Fourier transform will lead to the classical \textit{ringing} and associated artefacts. The robust removal of such discontinuities especially in the presence of noise and other fluctuations is not straightforward and is a subject of ongoing research \cite[see e.g.][]{gol88, sch03, vol03, kat08, she16}. Though we do not explore this variant approach in detail in this paper, we present an outline (see \S\ref{sec:bsp-angle-approach}) of the potential advantages using $\phi_\nabla^\textrm{m}$ could hold in comparison to using $e^{i\phi_\nabla^\textrm{m}}$.

\begin{figure}[htb]
\includegraphics[width=0.8\linewidth]{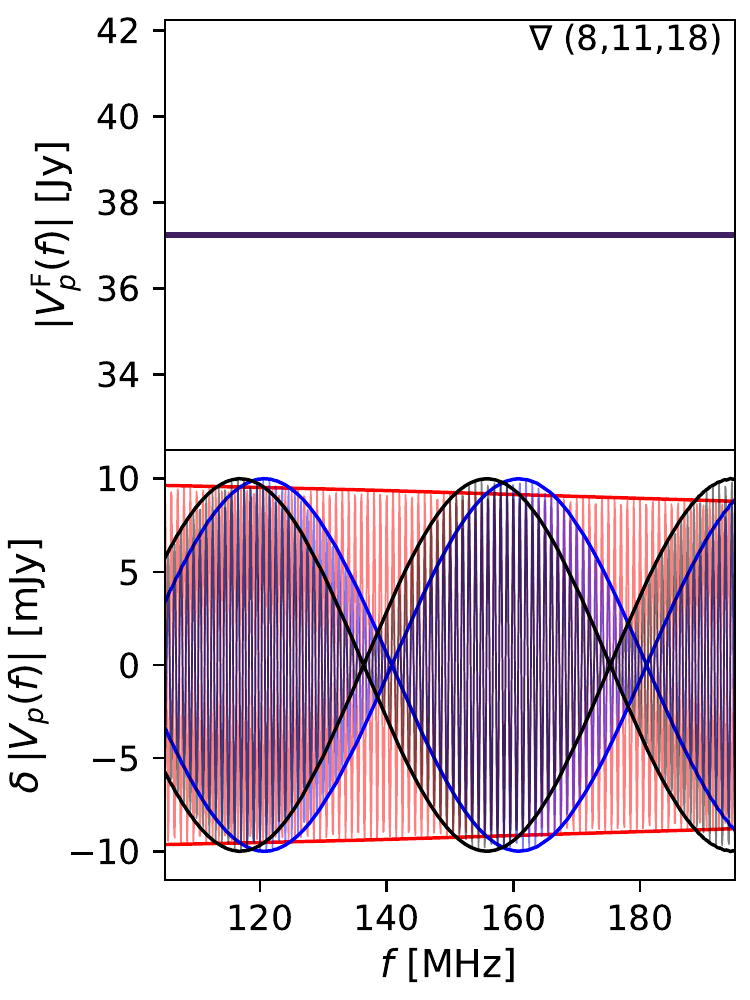}
\caption{The amplitude of the visibilities from the foregrounds, $V_p^\textrm{F}(f)$ (top) and the fluctuations therein due to the cosmic \HI\ signal (bottom) given by $\delta|V_p(f)|=|V_p^\textrm{F}(f)+V_p^\textrm{L}(f)|-|V_p^\textrm{F}(f)|$ on three antenna spacings (red, blue, and black) comprising the 50.6~m equilateral triad, $\nabla=(8,11,18)$ (refer to the antenna layout in Figure~\ref{fig:antenna-layout}). The foreground model is a point source of true flux density 100~Jy with spectral index, $\alpha=0$, located off-boresight by $\approx 5^\circ$. The power pattern at this location is the cause of the reduced strength of the apparent foreground visibilities in the top panel. The cosmic \HI\ signal is a purely hypothetical but an unrealistic model consisting of a point source located at boresight and has a cosine-shaped frequency spectrum of characteristic frequency scale, $\delta f_\textrm{L}=1/\tau_\textrm{L}=1$~MHz. The envelope of the fluctuations in amplitude obtained using a Hilbert transform is shown to gauge the overall magnitudes of these fluctuations relative to the foreground amplitudes. A color version of this figure is available in the online journal. \label{fig:vismodel-1ps-no-spindex}}
\end{figure}

Figure~\ref{fig:vismodel-1ps-no-spindex} corresponds to a hypothetical example that will be discussed in detail later in \S\ref{sec:1ps_FG_cosine_HI_displaced_spectrum} with a slight modification. Briefly, the foreground model is a point source of true flux density 100~Jy, which is $\approx 5^\circ$ off-boresight and has a spectral index $\alpha=0$ in contrast with the example in \S\ref{sec:1ps_FG_cosine_HI_displaced_spectrum}. The spectral signal comes from a point source at boresight with a cosine-shaped spectrum of characteristic frequency scale, $f_\textrm{L}=1$~MHz, and amplitude 10~mJy. Such a model for the foreground, and especially the EoR \HI\ signal, is purely hypothetical and unrealistic. The antenna spacings correspond to a 50.6~m equilateral triad (antenna layout discussed in \S\ref{sec:examples} and illustrated in Figure~\ref{fig:antenna-layout}). The antennas are assumed to be a uniformly illuminated dishes of diameter 14~m and have a corresponding \textit{Airy} angular power pattern whose transverse angular structure does not change with frequency. The top subpanel shows the visibility amplitude of the off-boresight foreground point source on the three differently oriented 50.6~m antenna spacings (red, blue, and black). The off-axis location lowers the apparent flux density as shown due to the power pattern assumed. The fluctuation in amplitude due to the spectral line point source, $\delta|V_p(f)|=|V_p^\textrm{F}(f)+V_p^\textrm{L}(f)|-|V_p^\textrm{F}(f)|$ for the same antenna spacings is shown in the bottom subpanel. The maximum of the envelope of the fluctuations in amplitude is $\sim 10$~mJy as expected. 

\begin{figure}[htb]
\includegraphics[width=0.8\linewidth]{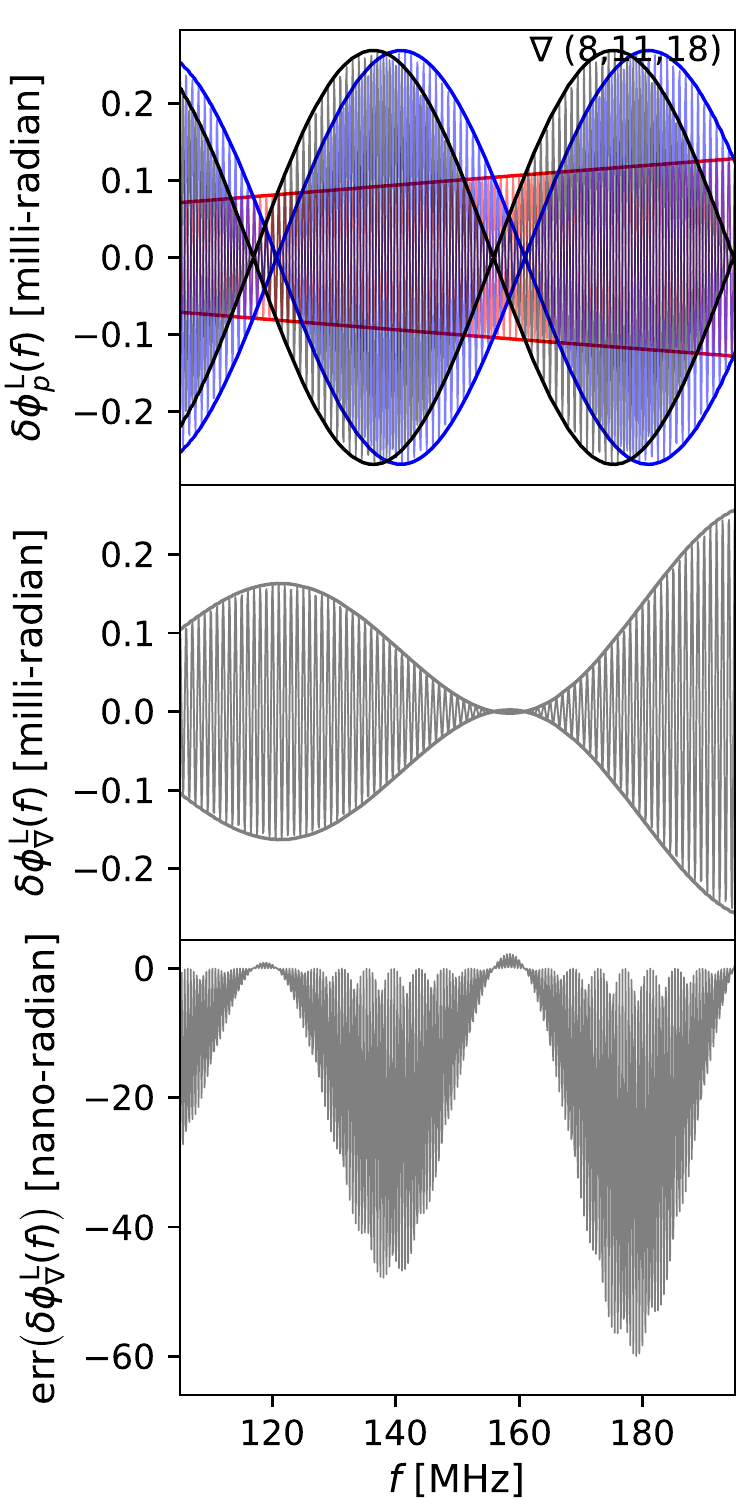}
\caption{Predicted and actual values of phase angle fluctuations of the visibility and bispectrum for the sky model considered in Figure~\ref{fig:vismodel-1ps-no-spindex}. \textit{Top:} Predicted fluctuations in the three visibility phase angles of a 50.6~m equilateral triad in red, blue, and black using Equation~(\ref{eqn:visphase-noise-perturbations}). \textit{Middle:} Fluctuations in the actual bispectrum phase angle (black) and the first-order approximation (gray) predicted using Equation~(\ref{eqn:bispectrum-line-perturbations-1}). The envelope of these phase angle fluctuations are at a level $\sim 0.1$--$0.2$~milli-radians, which is similar to and identifiable with the ratio $\sim \delta|V_p(f)|/|V_p^\textrm{F}(f)|$ in Figure~\ref{fig:vismodel-1ps-no-spindex}. \textit{Bottom:} The bispectrum phase angle residuals from the difference of the first-order prediction and the actual values. These higher-than-linear-order residuals not captured by the first-order approximation are $\sim$~few tens of nano-radians indicating the approximation to linear-order terms has a fractional inaccuracy of only $\lesssim 10^{-4}$ relative to true values. A color version of this figure is available in the online journal. \label{fig:closure_phase_comparison_no_spindex}}
\end{figure}

Figure~\ref{fig:closure_phase_comparison_no_spindex} shows the actual and predicted (from first-order approximation) values of the fluctuations in the phase angles of the visibilities and the bispectrum for the above example. The visibility phase angle fluctuations predicted after retaining only the first-order perturbations using Equation~(\ref{eqn:visphase-noise-perturbations}) for the three antenna spacings (red, blue, and black) are shown in the top subpanel. The middle subpanel shows the actual fluctuations in the bispectrum phase angle (black) and that predicted from Equation~(\ref{eqn:bispectrum-line-perturbations-1}) (gray). The envelope of the fluctuations derived using a Hilbert transform shows the overall amplitude of the phase angle fluctuations. The residuals between the predicted and actual bispectrum phase angle fluctuations is shown in the bottom subpanel. The higher-than-linear-order terms are typically $\sim 10^{-4}$ smaller fractionally compared to both the predicted linear-order terms as well as the actual values indicating that the fractional inaccuracy of the linear-order approximation is only $\lesssim 10^{-4}$ relative to true values. 

\begin{figure}[htb]
\includegraphics[width=0.8\linewidth]{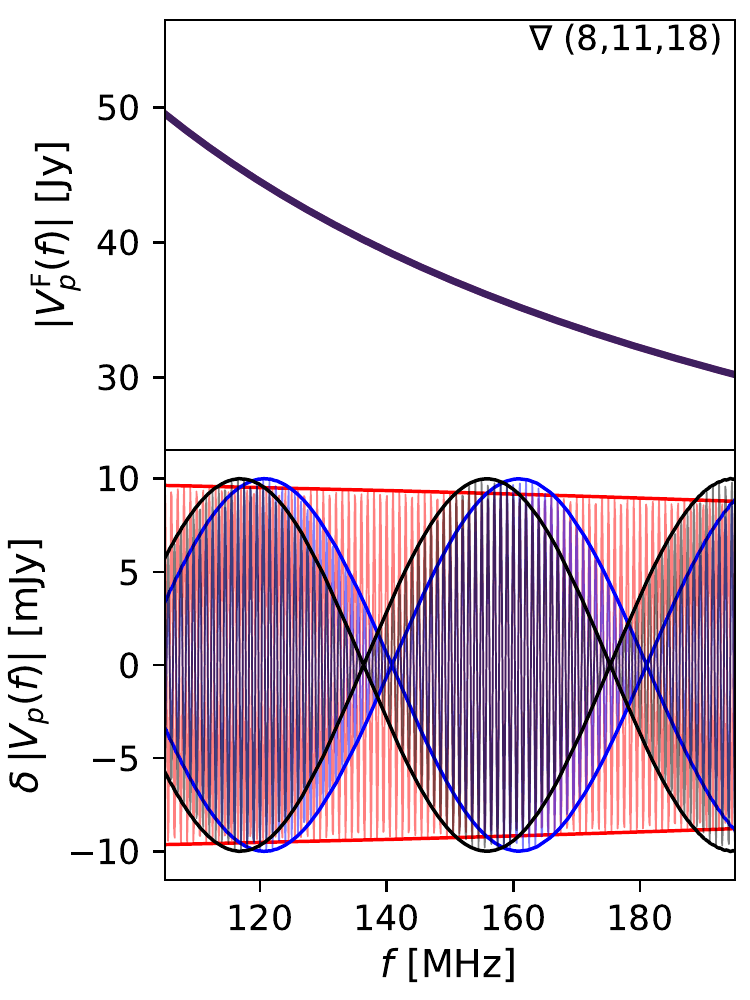}
\caption{Same as Figure~\ref{fig:vismodel-1ps-no-spindex} but the foreground model has a spectral index, $\alpha=-0.8$. A color version of this figure is available in the online journal. \label{fig:vismodel-1ps-spindex}}
\end{figure}

\begin{figure}[htb]
\includegraphics[width=0.8\linewidth]{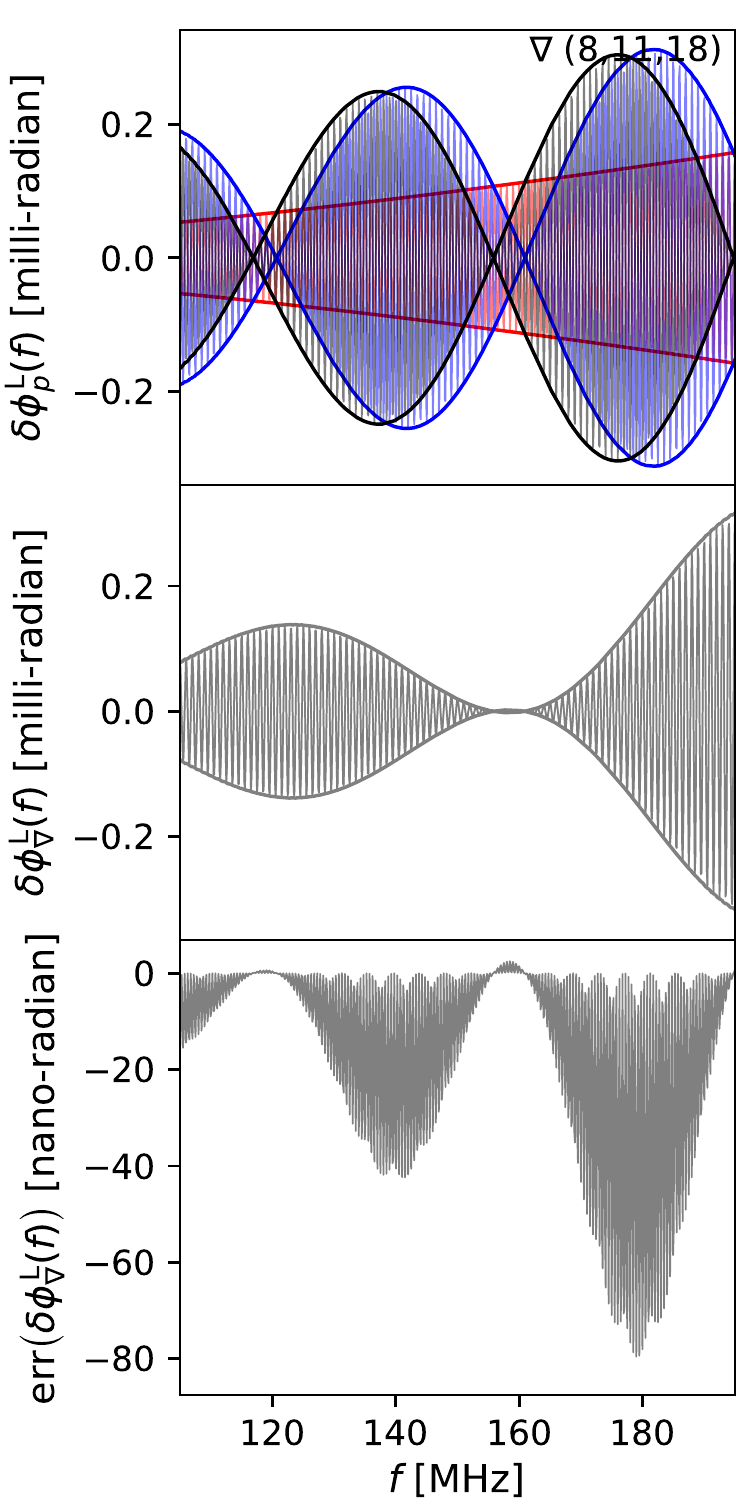}
\caption{Same as Figure~\ref{fig:closure_phase_comparison_no_spindex} but the foreground model has a spectral index, $\alpha=-0.8$ corresponding to that shown in Figure~\ref{fig:vismodel-1ps-spindex}. The visibility and bispectrum phase angle fluctuations are higher at higher frequencies and vice versa when compared to the case with $\alpha=0$ shown in Figure~\ref{fig:closure_phase_comparison_no_spindex}. This is because the fluctuations in visibility amplitudes remain the same whereas the foreground amplitudes are lower at higher frequencies and vice versa due to the spectral index, $\alpha<0$. Thus, the deviation between the actual and first-order approximation of the bispectrum phase also becomes higher at higher frequencies and vice versa relative to that when $\alpha=0$ in Figure~\ref{fig:closure_phase_comparison_no_spindex}. The approximation is still accurate overall, but is slightly better at lower frequencies than at higher frequencies essentially following the ratio $\sim \delta|V_p(f)|/|V_p^\textrm{F}(f)|$.A color version of this figure is available in the online journal. \label{fig:closure_phase_comparison_spindex}}
\end{figure}

Figures~\ref{fig:vismodel-1ps-spindex} and \ref{fig:closure_phase_comparison_spindex} are the same as Figures~\ref{fig:vismodel-1ps-no-spindex} and \ref{fig:closure_phase_comparison_no_spindex} respectively, but use a spectral index $\alpha=-0.8$ for the foreground point source. This same example is presented in more detail in \S\ref{sec:1ps_FG_cosine_HI_displaced_spectrum}. The phase angle fluctuations in visibility and bispectrum (Figure~\ref{fig:closure_phase_comparison_spindex}) are seen to be correspondingly increased at higher frequencies, and vice versa, relative to that in the previous example illustrated in Figure~\ref{fig:closure_phase_comparison_no_spindex}. This is also in agreement with the predictions in Eqs.~(\ref{eqn:visphase-noise-perturbations}) and (\ref{eqn:bispectrum-line-perturbations-1}) and arises due to the relative decrease of foreground amplitude at higher frequencies and vice versa. The deviation between the predicted and actual values also follows a similar trend where it is higher at higher frequencies and vice versa relative to the previous example. In other words, the first-order approximation is still valid at all frequencies but the prediction is slightly better at lower frequencies than at higher frequencies.

If a model of the foreground visibilities, $\widehat{V}_p^\textrm{F}$, is available, the spectral line strength from cosmic structures can be approximately estimated from this ratio. We construct the quantity
\begin{align}
V_\nabla(f) &= V_\textrm{eff}^\textrm{F}\,e^{i\phi_\nabla^\textrm{m}(f)}, \label{eqn:visscale}
\end{align}
where, $V_\textrm{eff}^\textrm{F}$ is designed to be an empirical estimate of $\widehat{V}_p^\textrm{F}(f)$. Note that the explicit dependence on $f$ has been re-introduced to emphasize that $V_\textrm{eff}^\textrm{F}$ is not a function of frequency but only provides an overall amplitude scaling that corresponds to the frequency band of interest. The multiplication by $V_\textrm{eff}^\textrm{F}$ that is empirically representative of the effective foreground visibility amplitude from the triads converts the complex bispectrum phase term to an effective flux density. $V_\nabla(f)$ has units of Jy but since it is purely mathematical rather than a physically valid flux density, we refer to its units by ``pseudo Jy'' to distinguish it from physical flux density. $V_\textrm{eff}^\textrm{F}$, a scalar, will not be able to fully capture the exact spatial coherence information in $V_p^\textrm{F}(f)$ because the latter consists of independent information on possibly three different spatial modes in addition to containing spectral information. This will introduce an error in the final result but one that appears as a simple scaling error and does not introduce any spectral errors or systematics.

Using Equation~(\ref{eqn:linearized-bispectrum-phase-visratio}), 
\begin{align}
    V_\nabla(f) &= V_\textrm{eff}^\textrm{F}\, e^{i\phi_\nabla^\textrm{F}(f)}\Biggl[1+\frac{1}{2}\sum_{p=1}^3\,\left(\frac{V_p^\textrm{L}(f)}{V_p^\textrm{F}(f)}-\frac{\conj{V_p^\textrm{L}}(f)}{\conj{V_p^\textrm{F}}(f)}\right) \nonumber \\ 
    &\qquad\qquad\qquad\quad + \frac{1}{2}\sum_{p=1}^3\,\left(\frac{V_p^\textrm{N}(f)}{V_p^\textrm{F}(f)}-\frac{\conj{V_p^\textrm{N}}(f)}{\conj{V_p^\textrm{F}}(f)}\right)\Biggr] \nonumber \\
    &= V_\nabla^\textrm{F}(f) + V_\nabla^\textrm{L}(f) + V_\nabla^\textrm{N}(f), \label{eqn:bsp-equivalence}
\end{align}
where, 
\begin{align}
    V_\nabla^\textrm{F}(f) &= V_\textrm{eff}^\textrm{F}\,e^{i\phi_\nabla^\textrm{F}(f)}, \\
    V_\nabla^\textrm{L}(f) &= i\,\delta\phi_\nabla^\textrm{L}(f)\,V_\textrm{eff}^\textrm{F}\, e^{i\phi_\nabla^\textrm{F}(f)}, \\
    \textrm{and,}\quad V_\nabla^\textrm{N}(f) &= i\,\delta\phi_\nabla^\textrm{N}(f)\,V_\textrm{eff}^\textrm{F}\, e^{i\phi_\nabla^\textrm{F}(f)}. 
\end{align}
Equation~(\ref{eqn:bsp-equivalence}) takes a familiar form where the measured visibilities comprise of an additive combination of foregrounds, the cosmological spectral line signal, and measurement noise. For convenience, we define $\gamma_p^\textrm{F}(f)=V_\textrm{eff}^\textrm{F}/V_p^\textrm{F}(f)$. Then, 
\begin{align}
    V_\nabla^\textrm{L}(f) &= \frac{1}{2}e^{i\phi_\nabla^\textrm{F}(f)}\sum_{p=1}^3\,\left[\gamma_p^\textrm{F}(f)V_p^\textrm{L}(f)-\conj{\gamma_p^\textrm{F}(f)V_p^\textrm{L}(f)}\right], \label{eqn:effective-line-visibilities} \\ 
    V_\nabla^\textrm{N}(f) &= \frac{1}{2}e^{i\phi_\nabla^\textrm{F}(f)}\sum_{p=1}^3\,\left[\gamma_p^\textrm{F}(f)V_p^\textrm{N}(f)-\conj{\gamma_p^\textrm{F}(f)V_p^\textrm{N}(f)}\right]. \label{eqn:effective-noise-visibilities}
\end{align}
However, differing from the standard approach, Eqs.~(\ref{eqn:effective-line-visibilities}) and (\ref{eqn:effective-noise-visibilities}) show that the effective visibilities denoting contributions of the spectral line signal and noise to the bispectrum phase are now weighted by the foreground spectra. This is expected because the bispectrum phase is a measure of the ratio of the fluctuating signal to the foregrounds as noted earlier. It must also be noted that $e^{i\phi_\nabla^\textrm{F}(f)}$, and $\gamma_p^\textrm{F}(f)$ (with $|\gamma_p^\textrm{F}(f)| \sim 1$) are expected to exhibit only slow spectral variations. Thus, the excess spectral variance from rapid fluctuations such as from the cosmic line signal will still be distinguishable as will be demonstrated later through examples.

The process of determining $V_\nabla$, as presented here, is empirical and has the following reasoning. We note that the variance due to cosmic line signal and noise fluctuations is approximately the sum of the variances in the fluctuations in the individual interferometric phases. Assuming that the cosmic line signal strength and the noise rms measured on each of the baselines forming the triad do not differ significantly between the baselines, the fluctuations are inversely dependent on the foreground visibility measured on the respective baselines. The weakest visibility amplitude among the baselines in the triad will induce the maximum fluctuations which will dominate the overall budget of fluctuations in the measured bispectrum phase. Therefore, we obtain $V_\textrm{eff}^\textrm{F}$ by averaging in inverse quadrature as:
\begin{align}
    \left(V_\textrm{eff}^\textrm{F}\right)^{-2} &= \sum_{p=1}^3 \, \left|\widehat{V}_p^\textrm{F}\right|^{-2}
\end{align}
where, 
\begin{align}
    \widehat{V}_p^\textrm{F} = \frac{\int \,  W(f)\,\widehat{V}_p^\textrm{F}(f)\,\mathrm{d}f}{\int W(f)\,\mathrm{d}f}.
\end{align}
$\widehat{V}_p^\textrm{F}(f)$ denotes a reliable visibility model (obtained either through calibration or modeling), which is then averaged over the frequency sub-band of interest, with the same optional spectral window weighting, $W(f)$, that may get used in the further processing as described below. It must be noted that the choice of $V_\textrm{eff}^\textrm{F}$ above is not entirely rigorous and could be substituted with any other reasonable estimate.

It must be emphasized that the model or calibrated $\widehat{V}_p^\textrm{F}(f)$ does not need to be fractionally as accurate as $\sim 10^{-5}$, for example, as in other standard approaches for detecting faint spectral lines. It is simply used to obtain an average scalar to scale the bispectrum phase to be in the same units as flux density. This procedure does not introduce any potential spectral artefact except for an overall uniform but minor error in the scaling because the choice of the scalar may not have been rigorous. 

\subsection{Delay Spectrum of the Bispectrum Phase}\label{sec:dspec-bsp}

Since the context of this paper is the detection of distinctive spectral features, we employ the delay spectrum technique \cite{par12a,par12b}, which is essentially a Fourier domain method for spectral discrimination. We define the delay transform, which is simply a Fourier transform, of a complex-valued spectrum, $Z(f)$, as:
\begin{align}\label{eqn:dspec}
  \widetilde{Z}(\tau) &= \int Z(f)\,e^{i2\pi f\tau}\,\mathrm{d}f.
\end{align}

Consider the delay-transform of $V_\nabla^\textrm{L}(f)$, denoted by $\widetilde{V}_\nabla^\textrm{L}(\tau)$. From Equation~(\ref{eqn:effective-line-visibilities}), it can be seen that $\widetilde{V}_\nabla^\textrm{L}(\tau)$ is formed from a convolution of $\widetilde{\mathcal{E}}_\nabla^\textrm{F}(\tau)$, $\widetilde{\gamma}_p^\textrm{F}(\tau)$, and $\widetilde{V}_p^\textrm{L}(\tau)$, which are the delay-domain duals of $e^{i\phi_\nabla^\textrm{F}(f)}$, $\gamma_p^\textrm{F}(f)$, and $V_p^\textrm{L}(f)$, respectively. $\widetilde{V}_p^\textrm{L}(\tau)$ contains the structural information about the cosmic spectral line signal. However, because of the weighting from foregrounds, it gets convolved by the predominantly smooth spectral structure response from the foregrounds within the sub-band in which the delay-transform is computed. 

\begin{align}\label{eqn:cpdspec-line}
  \widetilde{V}_\nabla^\textrm{L}(\tau) &= \frac{1}{2} \, \widetilde{\mathcal{E}}_\nabla^\textrm{F}(\tau)\ast \sum_{p=1}^3 \, \Bigl[\widetilde{\gamma}_p^\textrm{F}(\tau)\ast\widetilde{V}_p^\textrm{L}(\tau) \nonumber \\ 
  &\qquad\qquad\qquad\qquad -  \conj{\widetilde{\gamma}_p^\textrm{F}}(-\tau)\ast\conj{\widetilde{V}_p^\textrm{L}}(-\tau)\Bigr]
\end{align}

\begin{align}\label{eqn:cpdspec-full}
  \widetilde{\Psi}_\nabla(\tau) &= \widetilde{V}_\nabla(\tau)\ast\widetilde{W}(\tau) = \int V_\nabla(f)\,W(f)\,e^{i2\pi f\tau}\,\mathrm{d}f, \nonumber \\
  &= \widetilde{W}(\tau) \ast \left[\widetilde{V}_\nabla^\textrm{F}(\tau) + \widetilde{V}_\nabla^\textrm{L}(\tau) + \widetilde{V}_\nabla^\textrm{N}(\tau)\right],
\end{align}
which can be further expanded as:
\begin{widetext}
\begin{align}\label{eqn:cpdspec-full-expanded}
  \widetilde{\Psi}_\nabla(\tau) &= \widetilde{W}(\tau) \ast \widetilde{\mathcal{E}}_\nabla^\textrm{F}(\tau) \ast\Biggl\{V_\textrm{eff}^\textrm{F}\,\delta(\tau) + \frac{1}{2}\sum_{p=1}^3\,\Bigl[\widetilde{\gamma}_p^\textrm{F}(\tau)\ast \Bigl(\widetilde{V}_p^\textrm{L}(\tau) + \widetilde{V}_p^\textrm{N}(\tau) \Bigr) -  \conj{\widetilde{\gamma}_p^\textrm{F}}(-\tau)\ast \Bigl(\conj{\widetilde{V}_p^\textrm{L}}(-\tau) + \conj{\widetilde{V}_p^\textrm{N}}(-\tau) \Bigr) \Bigr]\Biggr\}
\end{align}
\end{widetext}
where, $W(f)$ is an optional spectral window weighting usually chosen to control the quality of the delay spectrum \citep{thy13,thy16} and has an effective bandwidth, $\Delta B$. $\widetilde{W}(\tau)$ is its delay-domain dual. $\delta(\tau)$ is a \textit{delta} function at $\tau=0$ in the \textit{delay} domain. $\widetilde{\Psi}_\nabla(\tau)$ has units of ``pseudo Jy~Hz'' for reasons explained earlier. 

\subsection{Delay Power Spectrum}\label{sec:pspec-bsp}

We obtain the analogous power spectrum of the bispectrum phase in the delay-domain as \citep{par12a,thy15a}:
\begin{align}
P_\nabla(\kappa_\parallel) &\equiv \bigl|\widetilde{\Psi}_\nabla(\tau)\bigr|^2 \left(\frac{A_\textrm{e}}{\lambda^2\Delta B}\right)\left(\frac{D^2\Delta D}{\Delta B}\right)\left(\frac{\lambda^2}{2k_\textrm{B}}\right)^2, \label{eqn:auto-delay-power-spectrum}
\end{align}
with
\begin{align}
  \kappa_\parallel &\equiv \frac{2\pi\tau\,f_\textrm{r}H_0\,E(z)}{c(1+z)^2}, 
\end{align}
where, $A_\textrm{e}$ is the effective area of the antenna, $\Delta B$ is the effective bandwidth, $\lambda$ is the wavelength of the band center, $k_\textrm{B}$ is the Boltzmann constant, $c$ is the speed of light in vacuum, $f_\textrm{r}$ is the rest-frame frequency of the cosmic spectral line signal, $z$ is the redshift, $D\equiv D(z)$ is the transverse comoving distance, and $\Delta D$ is the comoving depth along the line of sight corresponding to $\Delta B$ at redshift $z$. $H_0$, $h$, and $E(z)\equiv [\Omega_\textrm{M}(1+z)^3+\Omega_\textrm{k}(1+z)^2+\Omega_\Lambda]^{1/2}$ are standard terms in cosmology. In this paper, we use cosmological parameters from \cite{planck15xiii} with $H_0=100\,h$~km~s$^{-1}$~Mpc$^{-1}$. $P_\nabla(\kappa_\parallel)$ is in units of ``pseudo $\textrm{mK}^2\,(\textrm{Mpc}/h)^3\,$''. Note that we use $\kappa_\parallel$ to explicitly distinguish it from the line-of-sight wavenumber $k_\parallel$ as the two are very similarly defined mathematically but are not exactly related because the origin of fluctuations in the bispectrum phase are not identical to those in standard visibilities. $\kappa_\parallel$ has units of ``pseudo $h$~Mpc$^{-1}\,$''.

In a scenario that includes noise, the noise bias can be avoided by estimating the delay cross-power spectrum by replacing $\bigl|\widetilde{\Psi}_\nabla(\tau)\bigr|^2$ in Equation~(\ref{eqn:auto-delay-power-spectrum}) with $\Re\bigl\{\widetilde{\Psi}_\nabla(\tau)\,\conj{\widetilde{\Psi}_\nabla^\prime}(\tau)\bigr\}$, where, $\widetilde{\Psi}_\nabla^\prime(\tau)$ is another independent realization of $\widetilde{\Psi}_\nabla(\tau)$. The cross term, $\Re\{\widetilde{\Psi}_\nabla(\tau)\,\conj{\widetilde{\Psi}_\nabla^\prime}(\tau)\}$, serves the purpose of removing the noise bias or systematics in case of non-redundancy, etc. assuming the cosmic signal component remains fully correlated in both $\widetilde{\Psi}_\nabla(\tau)$ and $\widetilde{\Psi}_\nabla^\prime(\tau)$. In a noiseless and an ideal scenario, the cross-power spectrum reduces to the auto-power spectrum given in Equation~(\ref{eqn:auto-delay-power-spectrum}).

Table~\ref{tab:symbols} lists the symbols most relevant in this formalism along with their descriptions. Column 1 contains the symbols. Sub-columns in column 1 separated by a vertical delimiter denote Fourier domain duals of each other with the first sub-column representing the spectral (frequency) domain, and the second its Fourier-dual in the delay domain. The second column provides a brief description of the symbol(s). The third column points to the section in the text in which is the symbol is primarily introduced.

\begin{table*}
  \caption{Description of symbols.\label{tab:symbols}}
  \begin{ruledtabular}
    \begin{tabular}{c|cp{4.5in}lr}
      \multicolumn{2}{c}{Symbol} & Description & Section \\
      \toprule
      \multicolumn{2}{c}{$Z$, $\conj{Z}$} & A complex number and its conjugate & \S\ref{sec:complex-algebra} \\
      \multicolumn{2}{c}{$p$} & Index of antenna spacing (baseline) vector & \S\ref{sec:visphase-math} \\
      \multicolumn{2}{c}{$\boldsymbol{b}_p$} & Antenna spacing (baseline) vector indexed by $p$ & \S\ref{sec:visphase-math} \\
      \multicolumn{2}{c}{$f$} & Frequency & \S\ref{sec:visphase-math} \\
      \multicolumn{2}{c}{$\tau$} & Delay, the Fourier dual of frequency, $f$ & \S\ref{sec:dspec-bsp} \\
      \multicolumn{2}{c}{$\kappa_\parallel$} & Pseudo line-of-sight wavenumber modes corresponding to $\tau$ & \S\ref{sec:pspec-bsp} \\
      \midrule
      \multicolumn{2}{c}{$V_p^\textrm{T}(f)$} & \textit{True sky} visibility spectrum on $\boldsymbol{b}_p$ & \S\ref{sec:visphase-math} \\
      \multicolumn{2}{c}{$V_p^\textrm{m}(f)$} & \textit{Measured} visibility spectrum on $\boldsymbol{b}_p$ & \S\ref{sec:visphase-math} \\
      $V_p^\textrm{F}(f)$ & $\widetilde{V}_p^\textrm{F}(\tau)$ & True \textit{foreground} visibility spectrum and its delay transform on $\boldsymbol{b}_p$ & \S\ref{sec:visphase-math}, \S\ref{sec:dspec-bsp} \\
      $V_p^\textrm{L}(f)$ & $\widetilde{V}_p^\textrm{L}(\tau)$ & True \textit{spectral line} visibility spectrum and its delay transform on $\boldsymbol{b}_p$ & \S\ref{sec:visphase-math}, \S\ref{sec:dspec-bsp} \\
      $V_p^\textrm{N}(f)$ & $\widetilde{V}_p^\textrm{N}(\tau)$ & \textit{Noise} visibility spectrum and its delay transform on $\boldsymbol{b}_p$ & \S\ref{sec:visphase-math}, \S\ref{sec:dspec-bsp} \\
      \midrule
      \multicolumn{2}{c}{$\phi_p^\textrm{m}(f)$} & Spectrum of interferometric phase angle on \textit{measured} visibility, $V_p^\textrm{m}(f)$ & \S\ref{sec:visphase-math} \\
      \multicolumn{2}{c}{$\phi_p^\textrm{T}(f)$} & Spectrum of interferometric phase angle on \textit{true} visibility, $V_p^\textrm{T}(f)$ & \S\ref{sec:visphase-math} \\
      \multicolumn{2}{c}{$\phi_p^\textrm{F}(f)$} & Spectrum of interferometric phase angle on \textit{foreground} visibility, $V_p^\textrm{F}(f)$ & \S\ref{sec:visphase-math} \\
      \multicolumn{2}{c}{$\delta\phi_p^\textrm{L}(f)$} & Spectrum of perturbed interferometric phase angle due to $V_p^\textrm{L}(f)$ & \S\ref{sec:visphase-math} \\
      \multicolumn{2}{c}{$\delta\phi_p^\textrm{N}(f)$} & Spectrum of perturbed interferometric phase angle due to $V_p^\textrm{N}(f)$ & \S\ref{sec:visphase-math} \\
      \midrule
      \multicolumn{2}{c}{$\widehat{V}_p^\textrm{F}(f)$} & Model of \textit{foreground} visibility spectrum, $V_p^\textrm{F}(f)$ & \S\ref{sec:visphase-math} \\
      \multicolumn{2}{c}{$\widehat{V}_p^\textrm{m}(f)$} & Estimate of \textit{measured} visibility spectrum, $V_p^\textrm{m}(f)$, using $\phi_p^\textrm{m}(f)$ and $\widehat{V}_p^\textrm{F}(f)$ & \S\ref{sec:visphase-math} \\
      \multicolumn{2}{c}{$\widehat{V}_p^\textrm{L}(f)$} & Estimate of spectrum of \textit{spectral line} visibility, $V_p^\textrm{L}(f)$ & \S\ref{sec:visphase-math} \\
      \multicolumn{2}{c}{$\widehat{V}_p^\textrm{N}(f)$} & Estimate of \textit{noise} visibility spectrum, $V_p^\textrm{N}(f)$ & \S\ref{sec:visphase-math} \\
      \midrule
      \multicolumn{2}{c}{$B_\nabla^\textrm{m}(f)$} & Spectrum of \textit{measured} visibility bispectrum & \S\ref{sec:bsp-math} \\
      \multicolumn{2}{c}{$B_\nabla^\textrm{F}(f)$} & Spectrum of \textit{foreground} visibility bispectrum & \S\ref{sec:bsp-linear-perturbations} \\
      \multicolumn{2}{c}{$B_\nabla^\textrm{L}(f)$} & Spectrum of perturbation to $B_\nabla^\textrm{F}(f)$ due to the cosmic \textit{spectral line} signal & \S\ref{sec:bsp-linear-perturbations} \\
      \multicolumn{2}{c}{$B_\nabla^\textrm{N}(f)$} & Spectrum of perturbation to $B_\nabla^\textrm{F}(f)$ due to \textit{noise} & \S\ref{sec:bsp-linear-perturbations} \\
      \midrule
      \multicolumn{2}{c}{$\phi_\nabla^\textrm{m}(f)$} & Spectrum of phase angle on \textit{measured} bispectrum, $B_\nabla^\textrm{m}(f)$ & \S\ref{sec:bsp-math} \\
      \multicolumn{2}{c}{$\phi_\nabla^\textrm{F}(f)$} & Spectrum of phase angle on \textit{foreground} bispectrum, $B_\nabla^\textrm{F}(f)$ & \S\ref{sec:bsp-math} \\
      \multicolumn{2}{c}{$\delta\phi_\nabla^\textrm{L}(f)$} & Spectrum of perturbation to $\phi_\nabla^\textrm{F}(f)$ due to the cosmic \textit{spectral line} signal & \S\ref{sec:bsp-math} \\
      \multicolumn{2}{c}{$\delta\phi_\nabla^\textrm{N}(f)$} & Spectrum of perturbation to $\phi_\nabla^\textrm{F}(f)$ due to \textit{noise} & \S\ref{sec:bsp-math} \\
      \midrule
      \multicolumn{2}{c}{$V_\textrm{eff}^\textrm{F}$} & A scalar estimate of $\widehat{V}_p^\textrm{F}(f)$ obtained empirically over the sub-band & \S\ref{sec:bsp-vis-relation} \\
      $V_\nabla(f)$ & $\widetilde{V}_\nabla(\tau)$ & Representation of $e^{i\phi_\nabla^\textrm{m}(f)}$ in flux density units & \S\ref{sec:bsp-vis-relation}, \S\ref{sec:dspec-bsp} \\
      $V_\nabla^\textrm{F}(f)$ & $\widetilde{V}_\nabla^\textrm{F}(\tau)$ & Representation of $e^{i\phi_\nabla^\textrm{F}(f)}$ in flux density units and its delay transform & \S\ref{sec:bsp-vis-relation}, \S\ref{sec:dspec-bsp} \\
      $V_\nabla^\textrm{L}(f)$ & $\widetilde{V}_\nabla^\textrm{L}(\tau)$ & Representation of $e^{i\delta\phi_\nabla^\textrm{L}(f)}$ in flux density units and its delay transform & \S\ref{sec:bsp-vis-relation}, \S\ref{sec:dspec-bsp} \\
      $V_\nabla^\textrm{N}(f)$ & $\widetilde{V}_\nabla^\textrm{N}(\tau)$ & Representation of $e^{i\delta\phi_\nabla^\textrm{N}(f)}$ in flux density units and its delay transform & \S\ref{sec:bsp-vis-relation}, \S\ref{sec:dspec-bsp} \\
      \midrule
      $W(f)$ & $\widetilde{W}(\tau)$ & Spectral window function and its delay transform & \S\ref{sec:dspec-bsp} \\
      $\gamma_p^\textrm{F}(f)$ & $\widetilde{\gamma}_p^\textrm{F}(\tau)$ & Normalized true \textit{foreground} response in visibility and its delay transform on $\boldsymbol{b}_p$ & \S\ref{sec:bsp-vis-relation}, \S\ref{sec:dspec-bsp} \\
      $e^{i\phi_\nabla^\textrm{F}(f)}$ & $\widetilde{\mathcal{E}}_\nabla^\textrm{F}(\tau)$ & Complex Eulerian representation of $\phi_\nabla^\textrm{F}(f)$ and its delay transform & \S\ref{sec:bsp-math}, \S\ref{sec:dspec-bsp} \\
      $V_\nabla(f)W(f)$ & $\widetilde{\Psi}_\nabla(\tau)$ & Windowed $V_\nabla(f)$ and its delay transform & \S\ref{sec:dspec-bsp} \\
      \multicolumn{2}{c}{$P_\nabla(\kappa_\parallel)$} & Delay-domain \textit{power spectrum} of windowed bispectrum phase, $V_\nabla(f)W(f)$ & \S\ref{sec:pspec-bsp} \\
      
    \end{tabular}
  \end{ruledtabular}
\end{table*}

\section{Demonstration with Examples of Sky Models}\label{sec:examples}

For purposes of demonstration, we assume ideal cases without noise or other systematics. Only foregrounds and fluctuations from the cosmic spectral line signal will be considered. Hence, the noise terms can be ignored, and the use of auto-power spectrum will suffice. We consider the redshifted 21~cm signal from \HI\ from the EoR as our target cosmic signal, and foregrounds in the corresponding 100--200~MHz frequency band. Note that some of the examples where the foregrounds and the \HI\ signal from the EoR, especially the latter, are modeled as point sources are purely hypothetical and unrealistic. The purpose of such examples is to progressively build an intuition for the behavior of the bispectrum phase from simple to intermediate scenarios, eventually culminating in a more realistic example towards the end.

Four examples of sky models are considered, which are hereafter denoted as (\romannumeral 1), (\romannumeral 2), (\romannumeral 3), and (\romannumeral 4). In all these examples, the boresight points to RA(J2000)$=05^\textrm{h}\, 32^\textrm{m}\, 39\fs32$, Dec(J2000)$=-30\arcdeg\, 44\arcmin\, 05\farcs1$. In example~(\romannumeral 1), the foreground model is a point source of strength $V^\textrm{F}(f)=A(\hat{\boldsymbol{s}}_\textrm{F})\,V^\textrm{F}(f/f_\textrm{F})^\alpha\,e^{-i2\pi (f/c)\boldsymbol{b}_p\cdot\hat{\boldsymbol{s}}_\textrm{F}}$ with spectral index $\alpha$, and a pivot frequency for reference $f_\textrm{F}$, from a location $\hat{\boldsymbol{s}}_\textrm{F}$ where the angular power pattern of the antenna is given by $A(\hat{\boldsymbol{s}})$. The cosmic signal is also a point source of strength $V^\textrm{L}(f) = A(\hat{\boldsymbol{s}}_\textrm{L})\,V^\textrm{L}\cos{(2\pi f\tau_\textrm{L}+\theta_0)}\,e^{-i2\pi (f/c)\boldsymbol{b}_p\cdot\hat{\boldsymbol{s}}_\textrm{L}}$ which is a cosine-shaped spectral ripple of characteristic frequency scale $\delta f_\textrm{L}=1/\tau_\textrm{L}$, and an arbitrary angle $\theta_0$, at location $\hat{\boldsymbol{s}}_\textrm{L}$. Both the foreground and the spectral line signal appear as point sources in the transverse plane with $|V^\textrm{L}(f)| \ll |V^\textrm{F}(f)|$. Specifically, we adopt the values $V^\textrm{F} = 100$~Jy, $f_\textrm{F}=150$~MHz, $\alpha=-0.8$, $V^\textrm{L}=10$~mJy, and $\delta f_\textrm{L}=1/\tau_\textrm{L}=1$~MHz. This example is further sub-divided into two cases: (a) the location of the foreground and the cosmic signal point sources are colocated, $\hat{\boldsymbol{s}}_\textrm{F} = \hat{\boldsymbol{s}}_\textrm{L}$, and exactly at boresight, and (b) the foreground and the cosmic signal locations are not colocated, $\hat{\boldsymbol{s}}_\textrm{F} \ne \hat{\boldsymbol{s}}_\textrm{L}$, $\hat{\boldsymbol{s}}_\textrm{L}$ points to the boresight, and $\hat{\boldsymbol{s}}_\textrm{F}$ points to $\approx 5$\arcdeg off-boresight.

In example~(\romannumeral 2), the foreground consists of objects from the GLEAM catalog \cite{hur17} within a circle of 30$^\circ$ diameter around boresight. The cosmic \HI\ signal is modeled as a point source at boresight with a cosine-shaped spectrum, same as in the previous example. In example~(\romannumeral 3), the foreground consists of a point source of flux density 100~Jy at 150~MHz, \textbf{located at boresight} and spectral index $\alpha=-0.8$ for the foreground model (as in example~(\romannumeral 1)(a)). As our fiducial EoR \HI\ model, we use the {\sc faint galaxies} model \cite{mes16,gre17b} publicly available\footnote{\url{http://homepage.sns.it/mesinger/EOS.html}} from 21cmFAST simulations \cite{mes11} centered on boresight. In example~(\romannumeral 4), the foreground model consists of objects from the GLEAM catalog of radio sources (as in example~(\romannumeral 2)) and the fiducial 21cmFAST EoR \HI\ model from example~(\romannumeral 3).

The examples are briefly summarized in Table~\ref{tab:examples}. We reiterate that the hypothetical models of the foregrounds and the cosmic spectral line signal, especially the latter, being point sources in examples~(\romannumeral 1), (\romannumeral 2), and (\romannumeral 3) are unrealistic. However, having a point source for the sky model results in the  vanishing of the bispectrum phase angle which serves as a useful point of reference. And the cosine-shaped spectrum having a single characteristic frequency scale (correspondingly an impulse in the Fourier domain) serves the very useful purpose of understanding the response (also referred to as the \textit{impulse response} or the \textit{transfer function}) of the bispectrum phase statistic towards a single impulse input. Example~(\romannumeral 4) presents a realistic realization of both foregrounds and the cosmic EoR \HI\ signal.

\begin{table}
  \caption{Description of examples of sky models.\label{tab:examples}}
  \begin{ruledtabular}
    \begin{tabular}{>{\centering\arraybackslash}p{0.14\linewidth}|>{\centering\let\newline\\ \arraybackslash}p{0.4\linewidth}|>{\centering\arraybackslash}p{0.4\linewidth}}
      \multirow{2}{*}{Example} & \multicolumn{2}{c}{Sky Model} \\
      \cline{2-3}
                               & Foregrounds & Cosmic \HI\ Signal \\
      \midrule
      (\romannumeral 1)(a) & Point source\footnotemark[1] at boresight\footnotemark[2] of strength 100~Jy at 150~MHz and $\alpha=-0.8$ & Point source\footnotemark[1] at boresight with cosine-shaped spectrum of amplitude 10~mJy and characteristic frequency scale of 1~MHz \\
      \cline{2-3}
      (\romannumeral 1)(b) & Same as example~(\romannumeral 1)(a)\footnotemark[1] but $\approx 5$\arcdeg off-boresight & Same as example~(\romannumeral 1)(a)\footnotemark[1] \\
      \midrule
      (\romannumeral 2) & Objects from the GLEAM catalog within 15\arcdeg of boresight & Same as example~(\romannumeral 1)(a)\footnotemark[1] \\
      \midrule 
      (\romannumeral 3) & Same as example~(\romannumeral 1)(a)\footnotemark[1] & {\sc Faint galaxies} model from 21cmFAST simulations centered on boresight \\
      \midrule
      (\romannumeral 4) & Same as example~(\romannumeral 2) & Same as example~(\romannumeral 3) \\
    \end{tabular}
  \end{ruledtabular}
  \footnotetext[1]{A point source model for the foregrounds and the EoR \HI\ signal, especially the latter, is unrealistic and purely hypothetical.}
  \footnotetext[2]{Boresight points to $\textrm{RA(J2000)}=05^\textrm{h}\, 32^\textrm{m}\, 39\fs32$, $\textrm{Dec(J2000)}=-30\arcdeg\, 44\arcmin\, 05\farcs1$.}
\end{table}

\begin{figure}[htb]
\includegraphics[width=0.95\linewidth]{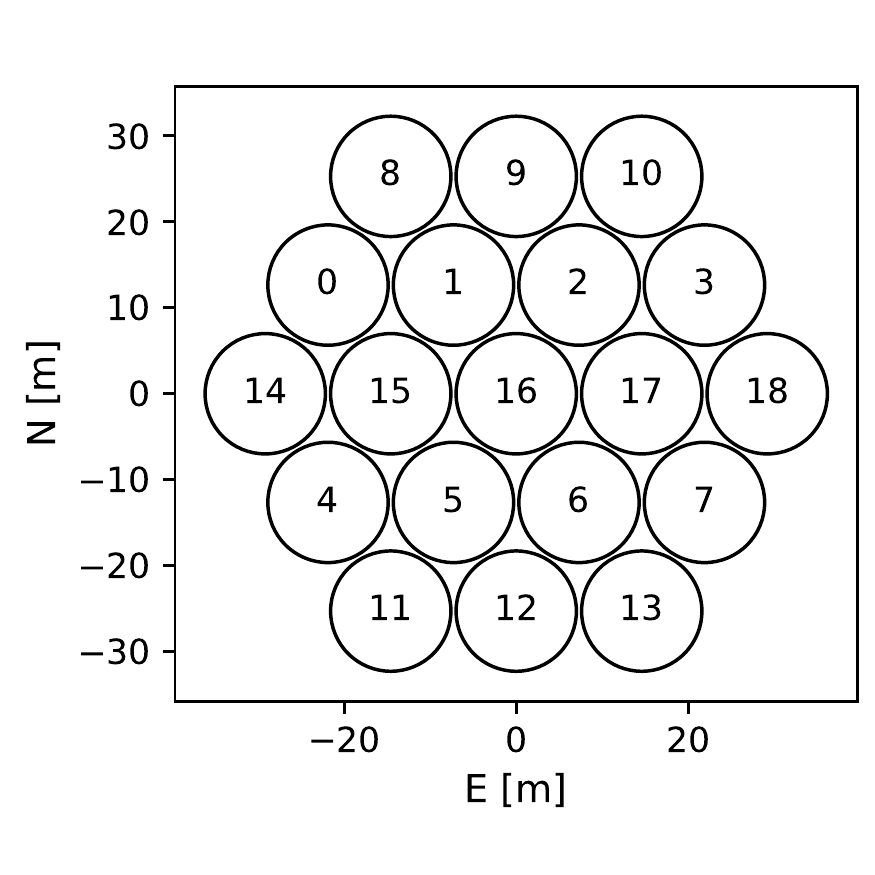}
\caption{Redundantly spaced antenna layout used in the simulations. They are assumed to be located at a latitude of $-30\fdg7224$ and a longitude of $+21\fdg4278$. The circles denote dish-shaped antennas, each of diameter 14~m. The numerals denote the antenna numbering. The $x$- and $y$-axes denote the local eastward and northward coordinates respectively. The array is assumed to be coplanar. The shortest spacing between antennas is 14.6~m. The specific triad classes chosen in this study are the 14.6~m (for example, $\nabla=(0,1,8)$) and 50.6~m equilateral triads (for example, $\nabla=(8,11,18)$). \label{fig:antenna-layout}}
\end{figure}

Figure~\ref{fig:antenna-layout} shows the antenna layout used in simulating visibilities, which is shown in local eastward and northward coordinates along the $x$- and $y$-axes respectively. The array is assumed to be coplanar and located at a latitude of $-30\fdg7224$ and a longitude of $+21\fdg4278$. The circles denote dish-shaped antennas each of diameter 14~m. The numerals denote the antenna numbering. The shortest antenna spacing is 14.6~m. The two classes of antenna triads frequently used in this paper are the 14.6~m and 50.6~m equilateral triads. Specific triads in each class are $\nabla=(0,1,8)$ and $\nabla=(8,11,18)$ respectively. 

Though the antenna placements are redundant, neither the mathematical formalism nor the results derived in this paper assume or require such a redundancy. The power patterns are also assumed to be identical between all the antennas with a uniform circular illumination of the aperture corresponding to an \textit{Airy} angular power pattern. In order to clearly isolate the findings reported in this paper from the spectral characteristics of the power pattern, we have further assumed that the power patterns are achromatic. The angular structure of the power pattern is identical at all frequencies in the 100--200~MHz band and is derived from the analytical expression evaluated at 150~MHz. 

We choose our spectral window function, $W(f)$ to be the ``modified'' Blackman-Harris window \cite{thy16} with an effective bandwidth of $\Delta B = 42$~MHz centered at 150~MHz. Although this choice of $\Delta B$ may include significant evolution of the properties of the cosmic signal within the sub-band, our aim in this paper is to demonstrate the spectral properties of bispectrum phase with highest resolution in Fourier space, $\delta\tau_\textrm{w}=1/\Delta B\approx 0.024\,\mu$s. This choice of $\Delta B$ is also the maximum that can fit inside the 100--200~MHz band without abrupt truncation of the window function at the edges. The effective area of the antennas is chosen to be $A_\textrm{e} = 100$~m$^2$ and is assumed to remain constant across the sub-band. Visibilities of the sky models in these examples were simulated using the Precision Radio Interferometry Simulator \cite[PRISim\footnote{\url{https://github.com/nithyanandan/PRISim}};][]{PRISim_software}.

\subsection{Frequency-domain view of the sky models}

The spectra of phases in the visibilities and the bispectrum are investigated to gain an understanding of their behavior in the frequency domain for each of the sky model examples. 

\subsubsection{Example~(\romannumeral 1): Unresolved Foreground and Unresolved Spectral Line Signal}\label{sec:1ps_FG_cosine_HI_spectrum}

From Equation~(\ref{eqn:visphase-line-perturbations}), the perturbation in the interferometric phase angle is given by
\begin{align}\label{eqn:HI-FG-PS-bsp}
    \delta\phi_p^\textrm{L}(f) &= \frac{A(\hat{\boldsymbol{s}}_\textrm{L})\,V^\textrm{L}\cos(2\pi f\tau_\textrm{L}+\theta_0)}{A(\hat{\boldsymbol{s}}_\textrm{F})\,V^\textrm{F}\left(\frac{f}{f_\textrm{F}}\right)^\alpha}\nonumber\\&\qquad\qquad\qquad\qquad\times \Im\left\{e^{-i 2\pi \frac{f}{c}  \boldsymbol{b}_p\cdot(\hat{\boldsymbol{s}}_\textrm{L} - \hat{\boldsymbol{s}}_\textrm{F})}\right\} \nonumber\\
    &= \frac{A(\hat{\boldsymbol{s}}_\textrm{L})\,V^\textrm{L}\cos(2\pi f\tau_\textrm{L}+\theta_0)}{A(\hat{\boldsymbol{s}}_\textrm{F})\,V^\textrm{F}\left(\frac{f}{f_\textrm{F}}\right)^\alpha} \nonumber\\
    &\qquad\qquad\qquad\times \sin\bigl(2\pi \frac{f}{c}  \boldsymbol{b}_p\cdot(\hat{\boldsymbol{s}}_\textrm{L} - \hat{\boldsymbol{s}}_\textrm{F})\bigr).
\end{align}
We define  $\tau_p(\hat{\boldsymbol{s}}_\textrm{L},\hat{\boldsymbol{s}}_\textrm{F})\equiv\tau_p\equiv \boldsymbol{b}_p\cdot(\hat{\boldsymbol{s}}_\textrm{L}-\hat{\boldsymbol{s}}_\textrm{F})/c$. 

For a point source foreground, regardless of $\alpha$, $\phi_\nabla^\textrm{F}(f) = 0$, $e^{i\phi_\nabla^\textrm{F}(f)} = 1$, and thus, $\widetilde{\mathcal{E}}_\nabla^\textrm{F}(\tau) = \delta(\tau)$. With $V_\textrm{eff}^\textrm{F}=V^\textrm{F}$, we get $\gamma^\textrm{F}(f)=(f/f_\textrm{F})^{-\alpha}$ which is a smooth function of frequency. Thus $\widetilde{\gamma}^\textrm{F}(\tau)$, the delay-domain dual of $\gamma^\textrm{F}(f)$, is a sharply peaked function in $\tau$. From Equation~(\ref{eqn:cpdspec-line}), the delay spectrum of the spectral line fluctuations in the bispectrum phase is given by:
\begin{align}\label{eqn:HI-FG-PS}
    \widetilde{V}_\nabla^\textrm{L}(\tau) &= \frac{A(\hat{\boldsymbol{s}}_\textrm{L})}{A(\hat{\boldsymbol{s}}_\textrm{F})}V^\textrm{L}\,\widetilde{W}(\tau)\ast\widetilde{\gamma}^\textrm{F}(\tau)\,e^{i\theta_0}\nonumber\\ 
    &\qquad\quad \ast\,\frac{1}{2}\left[\delta(\tau-\tau_\textrm{L}\right)+\delta\left(\tau+\tau_\textrm{L})\right] \nonumber\\
    &\qquad\quad \ast \frac{1}{2i}\sum_{p=1}^3\,[\delta(\tau-\tau_p)-\delta(\tau+\tau_p)] \nonumber\\
    &= \frac{A(\hat{\boldsymbol{s}}_\textrm{L})}{A(\hat{\boldsymbol{s}}_\textrm{F})}\frac{V^\textrm{L}\widetilde{W}(\tau)\ast\,\widetilde{\gamma}^\textrm{F}(\tau)\,e^{i\theta_0}}{4i}\nonumber\\
    &\quad \ast\sum_{p=1}^3\,\bigl[\delta\bigl(\tau-(\tau_p + \tau_\textrm{L})\bigr) + \delta\bigl(\tau-(\tau_p - \tau_\textrm{L})\bigr) \nonumber\\ 
    &\quad\,\,\,\,\, - \delta\bigl(\tau+(\tau_p + \tau_\textrm{L})\bigr) - \delta\bigl(\tau+(\tau_p - \tau_\textrm{L})\bigr)\bigr]. 
\end{align}
Without loss of generality, we can choose $\theta_0 = 0$. 

\paragraph{Spectral Line Signal Transversally Colocated with the Foreground}\label{sec:1ps_FG_cosine_HI_colocated_spectrum}

When the cosmic signal and the foreground object both modeled as point sources are colocated, $\hat{\boldsymbol{s}}_\textrm{F} = \hat{\boldsymbol{s}}_\textrm{L} = \hat{\boldsymbol{s}}_0$, then $\delta\phi_p^\textrm{L}(f) = 0$, and thus $\delta\phi_\nabla^\textrm{L}(f) = 0$. This can be qualitatively reasoned as follows. Regardless of the spectral structure, in any given frequency channel, the sky appears as a point source in the transverse sky plane. Therefore, $\phi_\nabla^\textrm{F}(f) = \delta\phi_\nabla^\textrm{L}(f) = 0$. This can be understood mathematically as well. The visibilities from the foregrounds and the cosmic spectral line signal are such that $\Im\{V_p^\textrm{L}(f)/V_p^\textrm{F}(f)\} = 0$. Therefore, from Equation~(\ref{eqn:bispectrum-line-perturbations-1}), $\delta\phi_\nabla^\textrm{L}(f) = 0$. This will also be true even without the linear-order approximation. Thus, even though the spectral structures are very different between the foreground and the hypothetical cosmic spectral line model, the spectral fluctuations from the latter will be indistinguishable from the foregrounds in $\phi_\nabla(f)$ because of the perfect relative symmetry in the transverse sky structure between the foregrounds and the cosmic signal (equivalent to the vanishing of the imaginary part in the visibility ratios).

\begin{figure}
\centering
  \subfloat[][Foreground and \HI\ visibility amplitudes \label{fig:EQ50_vis_1ps_FG_spindex_HI_colocated_achrAiry}]{\includegraphics[width=0.48\linewidth]{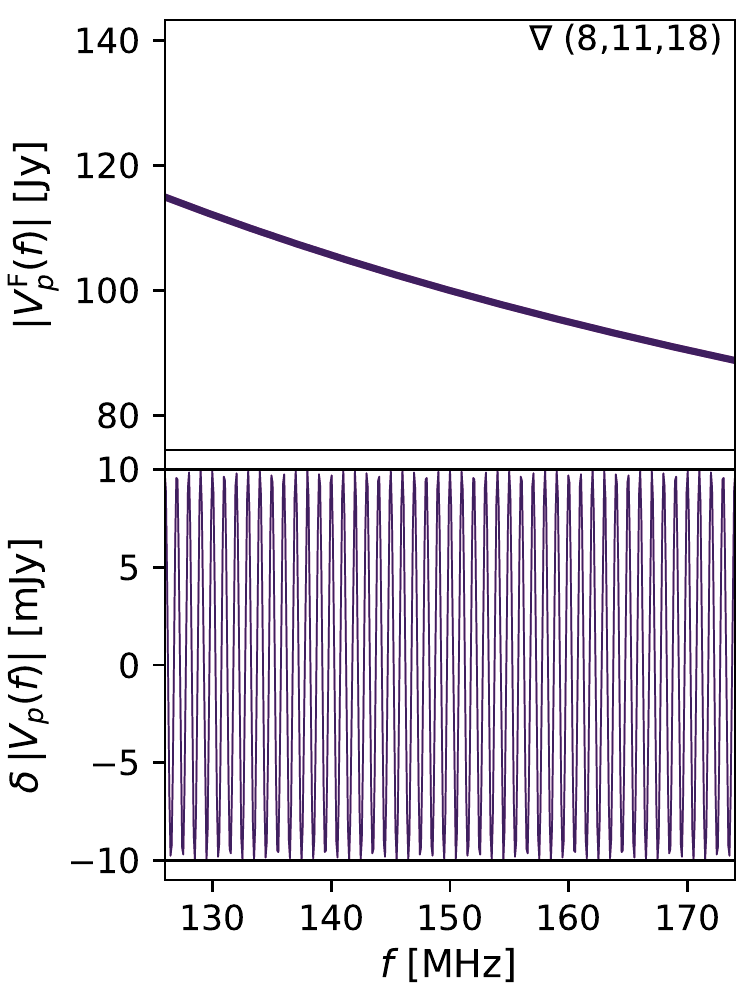}}
  \hspace{0.2cm}
  \subfloat[][Phase angle fluctuations in visibilities and bispectrum \label{fig:EQ50_phase_1ps_FG_spindex_HI_colocated_achrAiry}]{\includegraphics[width=0.48\linewidth]{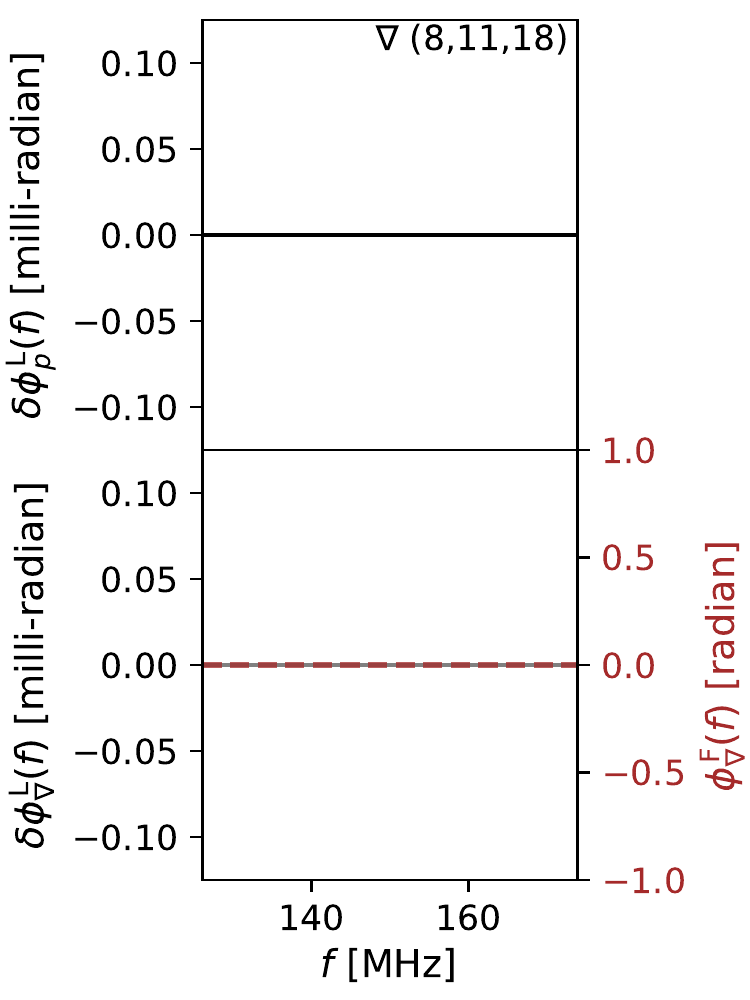}}
\caption{\textit{Left:} Foreground visibility amplitudes for a point source with a spectral index, $\alpha=-0.8$ (top) and the fluctuations therein (bottom), $\delta |V_p(f)| = |V_p^\textrm{F}(f)+V_p^\textrm{H}(f)| - |V_p^\textrm{F}(f)|$, for example~(\romannumeral 1)(a) measured on three antenna spacings (red, blue, and black) comprising the 50.6~m equilateral triad \textit{Right:} Perturbations due to the fluctuating \HI\ spectrum in the three visibility phase angles, $\delta\phi_p^\textrm{L}(f)$, in red, blue, and black (top) and the bispectrum phase angle (bottom) with the foreground component $\phi_\nabla^\textrm{F}(f)=0$ (dashed brown curve to be read off the $y$-axis placed on the right.) and the fluctuations caused by the cosmic \HI\ component $\delta\phi^\textrm{L}(f)$ (solid gray, $y$-axis on the left). Phase angle fluctuations in both the visibilities and the bispectrum are absent because of the colocation and symmetry in the transverse structure of the cosmic \HI\ signal relative to the point source foreground model. The frequency ranges in the $x$-axis are restricted only to enhance readability. A color version of this figure is available in the online journal. \label{fig:EQ50_vis_bsp_1ps_FG_spindex_HI_colocated_achrAiry}}
\end{figure}

Figure~\ref{fig:EQ50_vis_1ps_FG_spindex_HI_colocated_achrAiry} shows for example~(\romannumeral 1)(a) the amplitude of the visibilities on the three baselines comprising the 50.6~m equilateral triad due to the foreground (top), and the perturbations in visibility amplitudes (bottom) obtained as $\delta |V_p(f)| = |V_p^\textrm{F}(f)+V_p^\textrm{H}(f)| - |V_p^\textrm{F}(f)|$ due to the hypothetical \HI\ spectral line signal.  Figure~\ref{fig:EQ50_phase_1ps_FG_spindex_HI_colocated_achrAiry} shows the fluctuations in the phase angles of the visibility on the three antenna spacings, $\delta\phi_p^\textrm{L}(f)$ (top) and the bispectrum, $\delta\phi_\nabla^\textrm{L}(f)$ (bottom). They are both identically zero as expected even without any linear-order approximation.

Although this example is unrealistic and the spectral structure of \HI\ fluctuations does not manifest at all in the bispectrum phase in a manner useful towards its detection, it nevertheless reveals an important property of the spectrum of bispectrum phase fluctuations. If the transverse structure of the fluctuating signal is in perfect relative symmetry with respect to the underlying foreground transverse structures, then regardless of their inherently distinct spectral structures, no spectral signatures from the fluctuating cosmic spectral line signal will manifest in the bispectrum phase.

\begin{figure*}
  \centering
  \subfloat[][Example~(\romannumeral 1)(b) \label{fig:EQ50_vis_1ps_FG_spindex_HI_displaced_achrAiry}]{\includegraphics[width=0.24\linewidth]{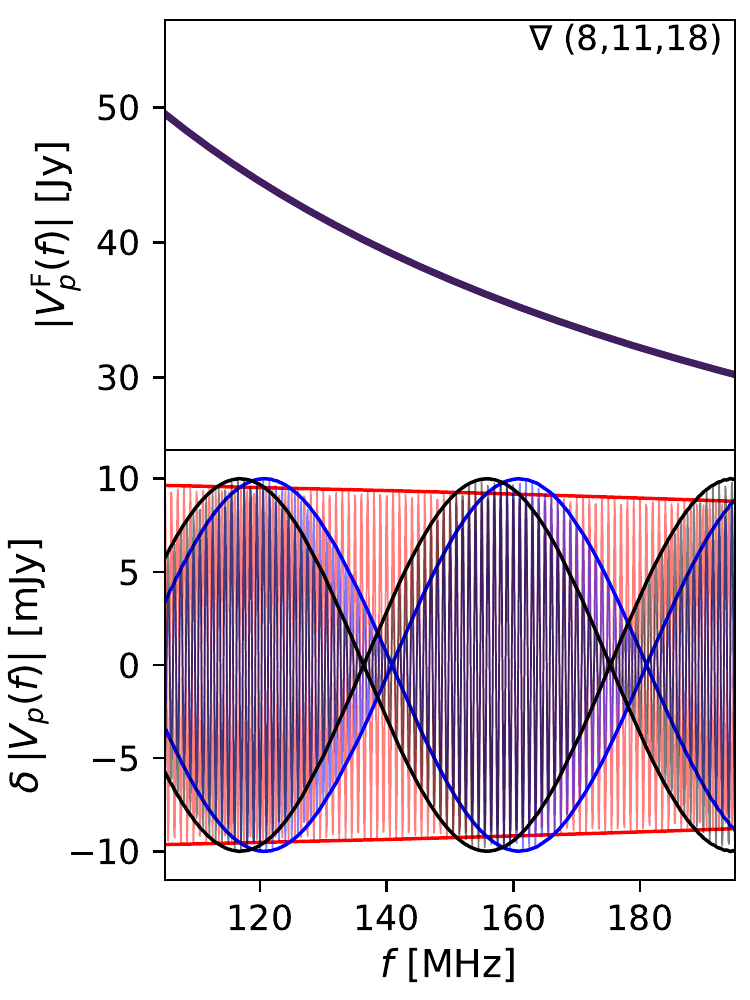}}
  \subfloat[][Example~(\romannumeral 2) \label{fig:EQ50_vis_GLEAM_FG_1ps_HI_achrAiry}]{\includegraphics[width=0.24\linewidth]{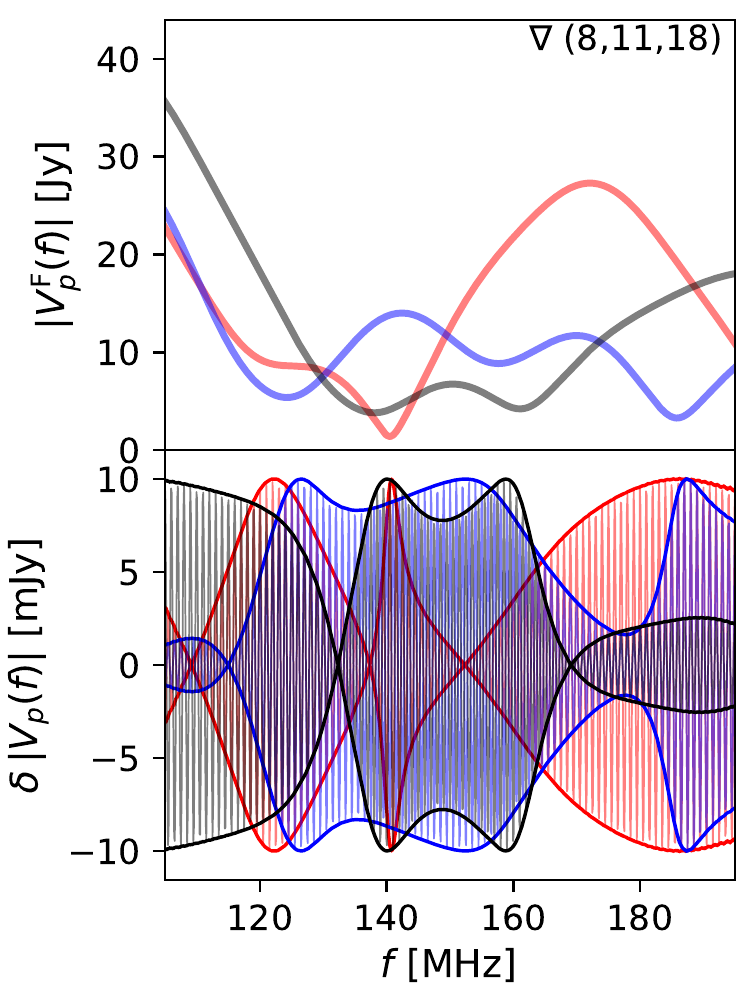}}
  \subfloat[][Example~(\romannumeral 3) \label{fig:EQ50_vis_1ps_FG_spindex_HI_21cmfast_achrAiry}]{\includegraphics[width=0.24\linewidth]{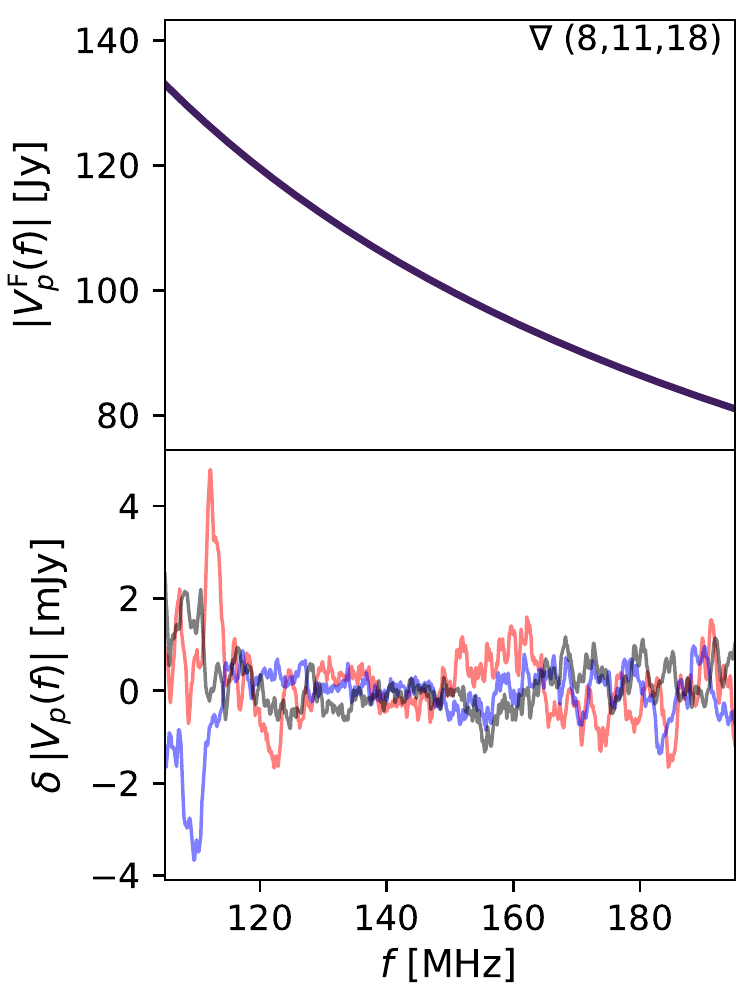}}
  \subfloat[][Example~(\romannumeral 4) \label{fig:EQ50_vis_GLEAM_FG_HI_21cmfast_achrAiry}]{\includegraphics[width=0.24\linewidth]{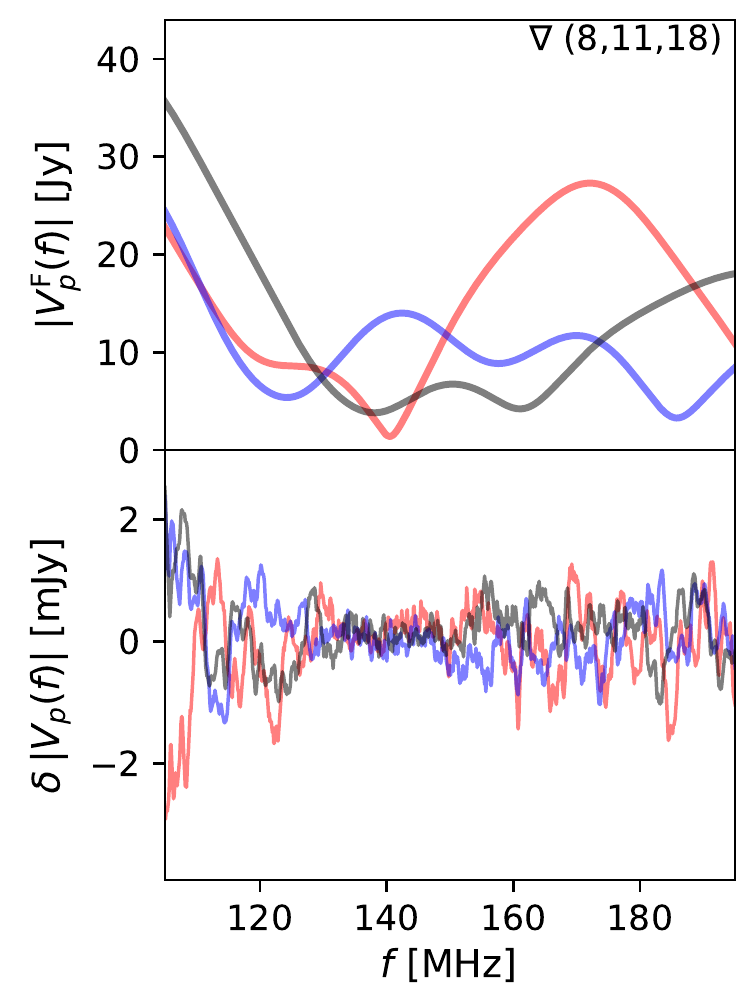}}
  \caption{Frequency spectra of amplitudes of visibilities due to foregrounds (top subpanels) and fluctuations therein caused by the cosmic \HI\ spectral line signal (bottom subpanels) for the examples specified. The 50.6~m equilateral triad used in these panels is specified at the top of the top subpanels. The three visibilities are shown in red, blue, and black. \textit{(a) Example~(\romannumeral 1)(b):} The foreground visibility amplitudes are smaller than in Figure~\ref{fig:EQ50_vis_1ps_FG_spindex_HI_colocated_achrAiry} by a factor equal to the power pattern of the \textit{Airy} disk at its angular separation from boresight. The visibility amplitude fluctuations are seen to have an envelope of amplitude $\sim 10$~mJy with the fastest spectral variation on scales of $\delta f_\textrm{L}=1/\tau_\textrm{L}=1$~MHz. The slower variation of the envelope is determined by the location of the cosmic \HI\ relative to the foreground point source. \textit{(b) Example~(\romannumeral 2):} The foreground visibilities obtained from the GLEAM catalog have a richer spectra owing to the wide-field distribution of foregrounds in the transverse direction, and yet exhibit smooth spectra. The broadband changes in the amplitude of the fluctuations are also due to the wide-field spatial distribution of the foreground objects. \textit{(c) Example~(\romannumeral 3):} The point source foreground has an extremely smooth spectrum, while the realistic \HI\ model exhibits rich spectral fluctuations on a wide range of scales. \textit{(d) Example~(\romannumeral 4):} Although the foregrounds from the GLEAM catalog show a rich foreground structure, they are still much smoother compared to the cosmic \HI\ spectral line signal obtained from the 21cmFAST simulations. A color version of this figure is available in the online journal. \label{fig:EQ50_vis_amp_spectra_examples}}
\end{figure*}

\begin{figure*}
  \centering
  \subfloat[][Example~(\romannumeral 1)(b) \label{fig:EQ50_phase_1ps_FG_spindex_HI_displaced_achrAiry}]{\includegraphics[width=0.24\linewidth]{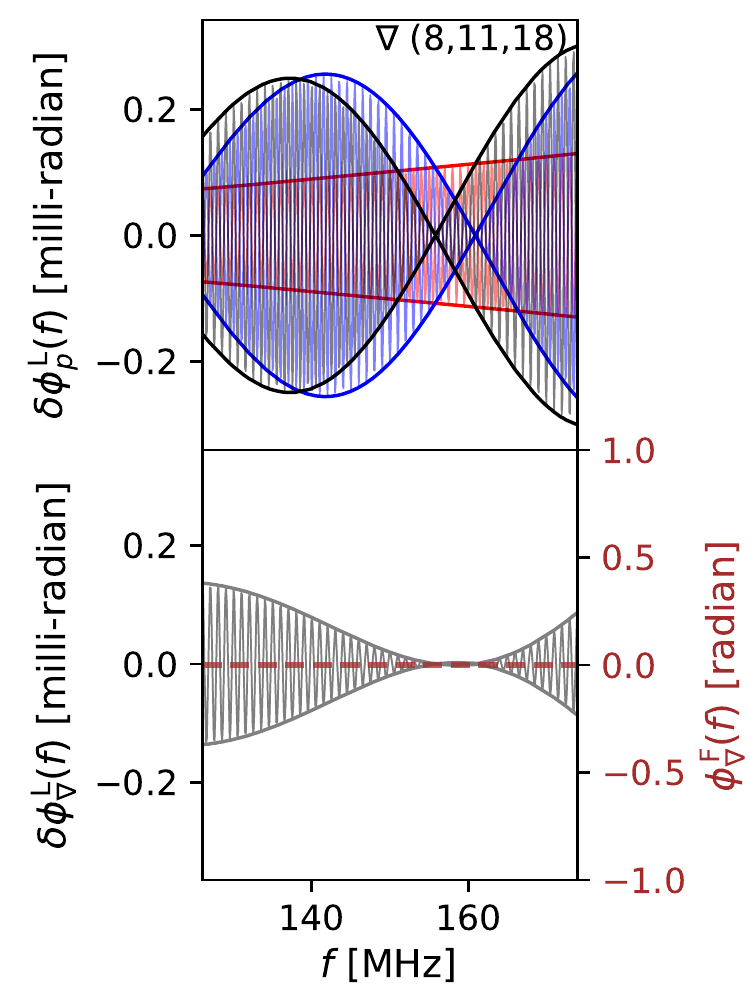}}
  \subfloat[][Example~(\romannumeral 2) \label{fig:EQ50_phase_GLEAM_FG_1ps_HI_achrAiry}]{\includegraphics[width=0.24\linewidth]{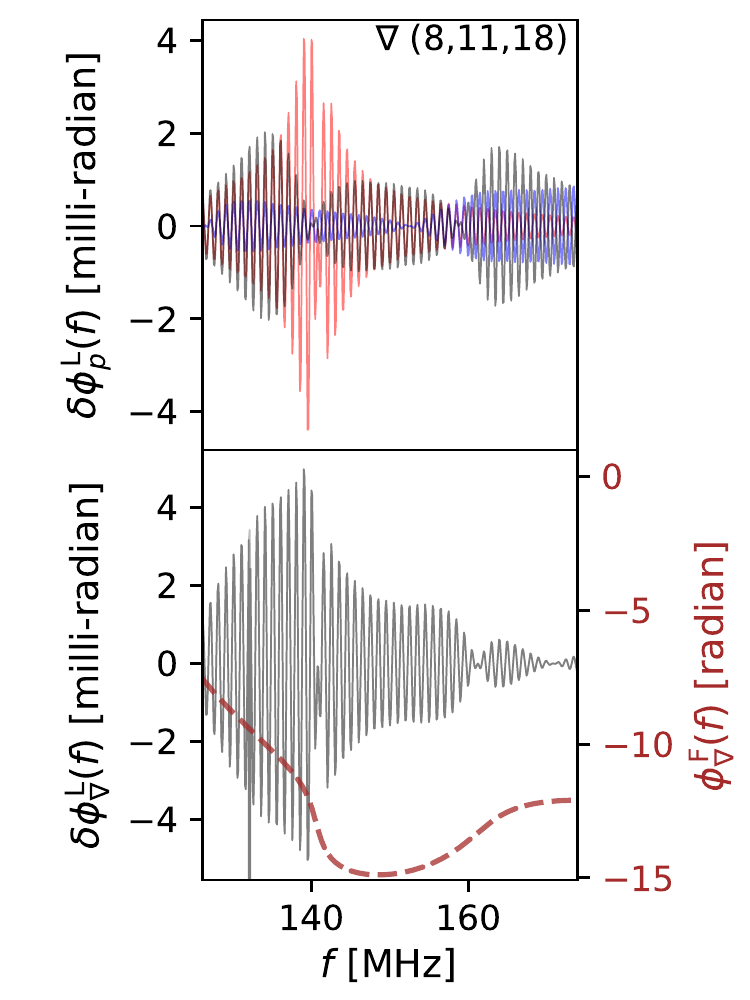}}
  \subfloat[][Example~(\romannumeral 3) \label{fig:EQ50_phase_1ps_FG_spindex_HI_21cmfast_achrAiry}]{\includegraphics[width=0.24\linewidth]{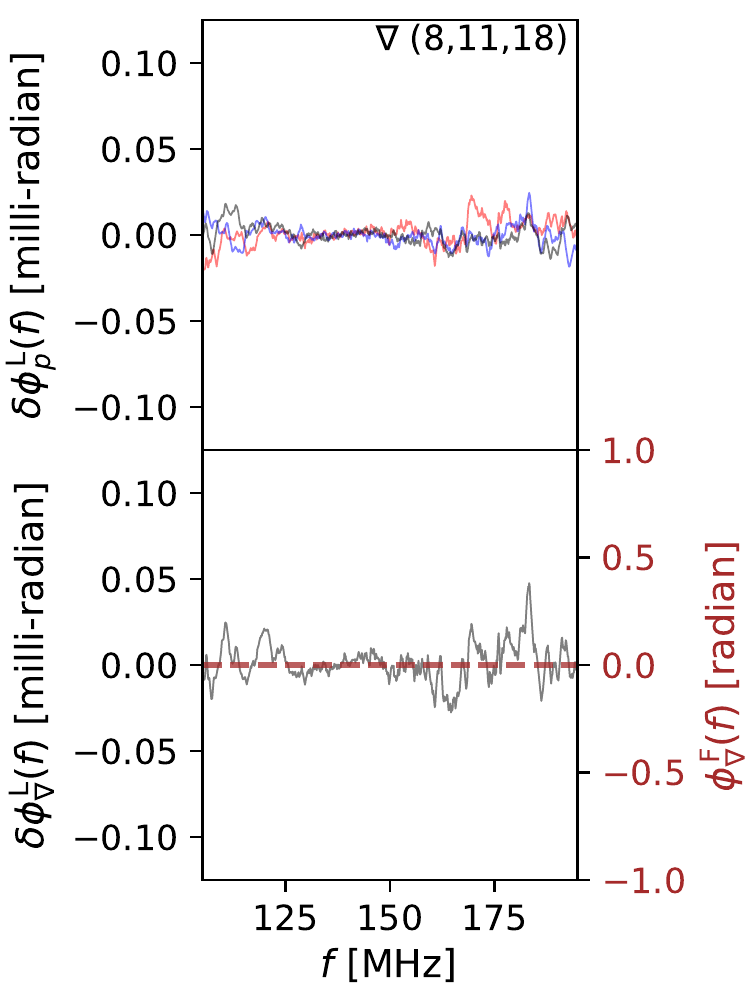}}
  \subfloat[][Example~(\romannumeral 4) \label{fig:EQ50_phase_GLEAM_FG_HI_21cmfast_achrAiry}]{\includegraphics[width=0.24\linewidth]{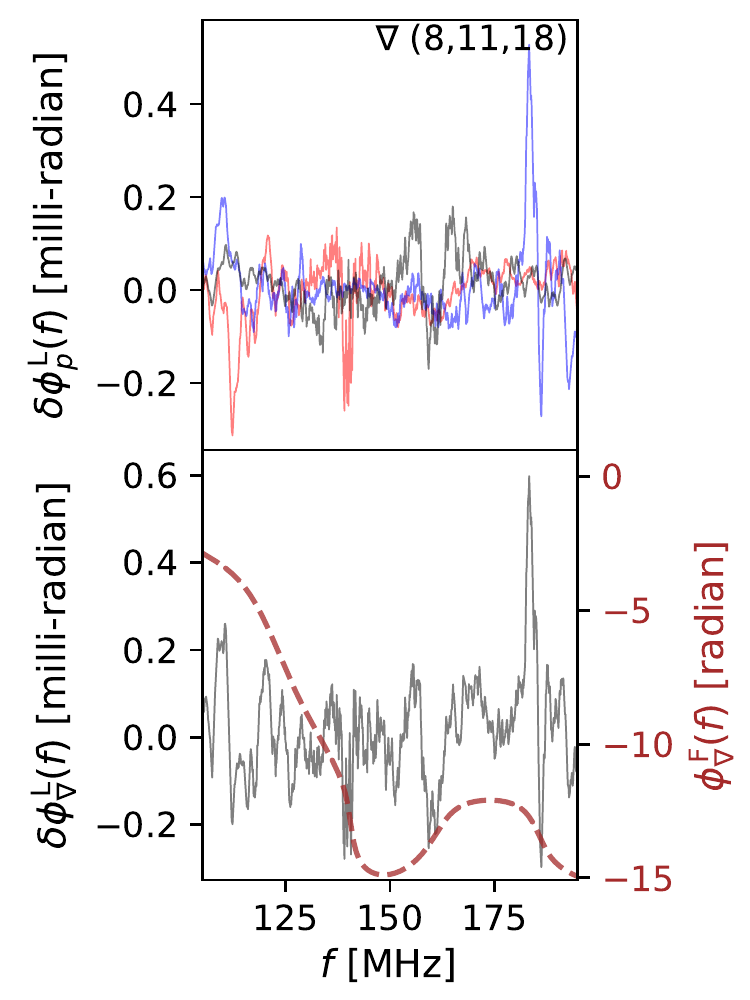}}
  \caption{The frequency spectra of the phase angle fluctuations on the visibilities (top subpanels) and the bispectrum due to the cosmic \HI\ spectral line fluctuations. The top subpanel shows the non-vanishing visibility phase fluctuations on the three antenna spacings (red, blue, and black). The bottom subpanels show the component of bispectrum phase from the foreground (dashed brown curve to be read off the $y$-axis placed on the right) and the bispectrum phase angle fluctuations (gray curves). The 50.6~m equilateral triad used in these panels is specified at the top of the top subpanels. \textit{(a) Example~(\romannumeral 1)(b):} As the foreground visibility amplitude decreases at higher frequencies due to the spectral index (see Figure~\ref{fig:EQ50_vis_1ps_FG_spindex_HI_displaced_achrAiry}), the amplitude of the phase angle fluctuations increases. Both these phase angle fluctuations are non-zero unlike when the cosmic \HI\ and foreground models were perfectly colocated with respect to each other in example~(\romannumeral 1)(a) (see Figure~\ref{fig:EQ50_phase_1ps_FG_spindex_HI_colocated_achrAiry}). \textit{(b) Example~(\romannumeral 2):} The foreground component of the bispectrum phase angle is non-zero and has smooth spectral structure which is unwrapped to remove discontinuities at odd multiples of $\pm\pi$. The envelope of phase fluctuations in both the visibilities (top subpanel) and the bispectrum (bottom subpanel) has amplitudes inversely proportional to the foreground visibility amplitudes (see Figure~\ref{fig:EQ50_vis_GLEAM_FG_1ps_HI_achrAiry}). \textit{(c) Example~(\romannumeral 3):} The point source foreground yields $\phi_\nabla^\textrm{F}(f) = 0$ (dashed brown curve with its $y$-axis placed on the right). There is a significant correspondence between the shape and scale of these phase angle fluctuations in visibilities and the bispectrum to those in the visibility amplitudes in the bottom subpanel of Figure~\ref{fig:EQ50_vis_1ps_FG_spindex_HI_21cmfast_achrAiry}. \textit{(d) Example~(\romannumeral 4):} The cosmic \HI\ spectral line signal from the 21cmFAST simulations shows a lot more spectral structure in the phase angle fluctuations of the visibilities and the bispectrum relative to the smooth spectral structure from the GLEAM foregrounds. There is a significant correspondence between the phase angle fluctuations shown here and the amplitude fluctuations in Figure~\ref{fig:EQ50_vis_GLEAM_FG_HI_21cmfast_achrAiry}. The sharp spikes in the spectra of phase angle fluctuations are generally regions with low foreground amplitudes (for example, near 140~MHz and 187~MHz). The frequency ranges in the $x$-axis are restricted only to enhance readability. A color version of this figure is available in the online journal. \label{fig:EQ50_phase_spectra_examples}}
\end{figure*}

Next, the frequency-domain behavior of the sky model examples~(\romannumeral 1)(b), (\romannumeral 2), (\romannumeral 3), and (\romannumeral 4) are discussed below in detail and illustrated collectively in Figures~\ref{fig:EQ50_vis_amp_spectra_examples} and \ref{fig:EQ50_phase_spectra_examples}, which characterize the amplitudes and phases, respectively.

\paragraph{Spectral Line Signal Transversally Displaced from the Foreground} \label{sec:1ps_FG_cosine_HI_displaced_spectrum}

In contrast to example~(\romannumeral 1)(a) in section~\ref{sec:1ps_FG_cosine_HI_colocated_spectrum}, usually $\exists\, \boldsymbol{b}_p$ such that $\boldsymbol{b}_p\cdot(\hat{\boldsymbol{s}}_\textrm{L} - \hat{\boldsymbol{s}}_\textrm{F}) = c\tau_p \ne 0$. Under such conditions, Eqs.~(\ref{eqn:HI-FG-PS-bsp}) and (\ref{eqn:HI-FG-PS}) yield a non-zero response, at a minimum of 4 and possibly up to 12 sharply-peaked distinct delays (four for each $p$) at $\tau = \pm (\tau_p - \tau_\textrm{L})$ and $\tau = \pm (\tau_p + \tau_\textrm{L})$. The delay spectrum of spectral line visibility can be expressed as delta functions at $\tau = \frac{\boldsymbol{b}_p\cdot \hat{\boldsymbol{s}}_\textrm{L}}{c}\pm\tau_\textrm{L}$:
\begin{align}
    \widetilde{V}_p^\textrm{L}(\tau) &= \frac{A(\hat{\boldsymbol{s}}_\textrm{L}) V^\textrm{L}e^{i\theta_0}}{2}\left[\delta(\tau-\tau_\textrm{L}\right)+\delta\left(\tau+\tau_\textrm{L})\right] \nonumber\\
                                    &\qquad\qquad\qquad\qquad \ast \delta(\tau-\frac{\boldsymbol{b}_p\cdot \hat{\boldsymbol{s}}_\textrm{L}}{c}) \ast \widetilde{W}(\tau) \nonumber\\
    &= \frac{A(\hat{\boldsymbol{s}}_\textrm{L})V^\textrm{L}e^{i\theta_0}}{2}\Bigl[\widetilde{W}\Bigl(\tau-(\frac{\boldsymbol{b}_p\cdot \hat{\boldsymbol{s}}_\textrm{L}}{c}+\tau_\textrm{L})\Bigr) \nonumber\\
    &\qquad\qquad\qquad +\widetilde{W}\Bigl(\tau-(\frac{\boldsymbol{b}_p\cdot \hat{\boldsymbol{s}}_\textrm{L}}{c}-\tau_\textrm{L})\Bigr)\Bigr].
\end{align}

Mathematically, the delay spectrum of the bispectrum phase using the linear approximation is expected to differ from the standard delay spectrum in the following ways:
\begin{enumerate}
    \item The former is sharply-peaked but broader than the latter because of the convolution with the delay response of the foreground spectrum term, $\widetilde{\gamma}^\textrm{F}(\tau)$. This broadening behavior is more clearly illustrated in other examples that follow.
    \item The former peaks at four distinct delays corresponding for each antenna spacing because the bispectrum phase fluctuations are a product of the cosine-shaped spectrum with the sine of the phase angle from the displacement of the cosmic \HI\ source relative to the foreground object, whereas the latter has only a pair of peaks for each visibility from the cosine-shaped spectrum. 
    \item The spatial position (and the spatial structure, in general) of the foreground influences where the delays are in the bispectrum phase delay spectrum. This reaffirms that the bispectrum phase fluctuations are a measure of the relative transverse-plane asymmetry (or dissimilarity) between the cosmic fluctuations and the dominant foregrounds, whose magnitude depends on the ratio between the two as predicted to first order by Eqs.~(\ref{eqn:bispectrum-line-perturbations-1}) and (\ref{eqn:bispectrum-line-perturbations-2}).  
\end{enumerate}

Figure~\ref{fig:EQ50_vis_1ps_FG_spindex_HI_displaced_achrAiry} shows that the positional displacement between the \HI\ and the foreground models causes a slow variation in the envelope of the visibility fluctuations with a maximum amplitude of 10~mJy and fastest spectral variations are on scales of $\delta f_\textrm{L}=1/\tau_\textrm{L}=1$~MHz.  Figure~\ref{fig:EQ50_phase_1ps_FG_spindex_HI_displaced_achrAiry} shows the fluctuations in the phase angles of the visibility (top subpanel) and the bispectrum (bottom subpanel). In contrast with example~(\romannumeral 1)(a) (see Figure~\ref{fig:EQ50_vis_1ps_FG_spindex_HI_colocated_achrAiry}), the phase angle fluctuations in both the visibilities and bispectrum phase are non-zero because of the relative asymmetry in the transverse structure between the foreground and the \HI\ model, and increase in amplitude towards higher frequencies due to the lowered foreground visibility amplitudes caused by $\alpha<0$. 

\subsubsection{Example~(\romannumeral 2): Realistic Foreground and Unresolved Spectral Line Signal}\label{sec:GLEAM_FG_cosine_HI_spectrum}

Here, we consider a foreground model determined by the GLEAM catalog \cite{hur17}. Visibilities were modeled for sources in the GLEAM catalog within a circle of 30$^\circ$ diameter around boresight. The cosmic \HI\ signal is modeled as the same unrealistic point source at boresight with a cosine-shaped spectrum as in the previous example.

Figure~\ref{fig:EQ50_vis_GLEAM_FG_1ps_HI_achrAiry} shows the amplitude of the foreground visibilities and the fluctuations in the amplitude due to the \HI\ fluctuations. The fastest spectral fluctuations are on scales of $\delta f_\textrm{L}=1/\tau_\textrm{L}=1$~MHz while the broadband fluctuations in the amplitude are determined by the locations of the various GLEAM catalog objects relative to the \HI\ point source at boresight. Figure~\ref{fig:EQ50_phase_GLEAM_FG_1ps_HI_achrAiry} shows the phase angle fluctuations in the visibility (top) and the bispectrum (bottom). The broadband changes in the amplitude of the phase angle fluctuations are due to the changing amplitudes of the foreground visibility spectra. Although the foregrounds have richer spectral structure they are still relatively smooth.

\subsubsection{Example~(\romannumeral 3): Unresolved Foreground and Fiducial Cosmic Spectral Signal} \label{sec:1ps_FG_21cmfast_HI_spectrum}

This example consists of a point source of flux density 100~Jy at 150~MHz, located at boresight and spectral index $\alpha=-0.8$ for the foreground model. As our fiducial EoR \HI\ model, we use the {\sc faint galaxies} model \cite{mes16,gre17b} from 21cmFAST simulations \cite{mes11} centered on boresight. 

Figure~\ref{fig:EQ50_vis_1ps_FG_spindex_HI_21cmfast_achrAiry} shows the foreground visibility amplitude (top subpanel) and the fluctuations in the amplitude caused by the cosmic \HI\ fluctuations (bottom subpanel). Figure~\ref{fig:EQ50_phase_1ps_FG_spindex_HI_21cmfast_achrAiry} shows the phase angle fluctuations in the visibilities (top subpanel) and the bispectrum (bottom subpanel). The phase angle fluctuations span a wide range of frequency scales. They are seen to be at a level $\lesssim 10^{-4}$ radians (which depends on the foreground visibility amplitudes) that is consistent with Eqs.~(\ref{eqn:bispectrum-line-perturbations-1}) and (\ref{eqn:bispectrum-line-perturbations-2}), and have coherent structures on frequency scales that approximately correspond to those in Figure~\ref{fig:EQ50_vis_1ps_FG_spindex_HI_21cmfast_achrAiry}. 

\subsubsection{Example~(\romannumeral 4): Realistic Foreground and Fiducial Cosmic Spectral Line Signal}\label{sec:GLEAM_FG_21cmfast_HI_spectrum}

We consider the GLEAM catalog of radio sources for our foreground model and the fiducial 21cmFAST EoR \HI\ model from the example above. This represents a realistic realization of both foregrounds and the cosmic EoR \HI\ signal. Figure~\ref{fig:EQ50_vis_GLEAM_FG_HI_21cmfast_achrAiry} shows the foreground visibility amplitudes and the fluctuations therein for a 50.6~m equilateral triad. Figure~\ref{fig:EQ50_phase_GLEAM_FG_HI_21cmfast_achrAiry} shows the phase angle fluctuations of the visibilities and the bispectrum. The sharp spikes, for example at $\simeq 186$~MHz, are due to low foreground amplitudes and thus reaffirm that they are a function of the ratio of the \HI\ fluctuations to the foreground strength. Although the foreground model from the GLEAM catalog exhibits a relatively rich spectral structure, they are still much smoother compared to the cosmic \HI\ spectral line fluctuations from the EoR obtained using 21cmFAST simulations. Redshifted 21~cm interferometer experiments aim to detect this spectral distinction.

\subsection{Delay- (Fourier-) domain view of the sky models}\label{sec:dspec}

The Fourier-domain view into the visibilities and the bispectrum phases of these examples are explored using their delay spectra.

\subsubsection{Example~(\romannumeral 1): Unresolved Foreground and Unresolved Spectral Line Signal}\label{sec:1ps_FG_cosine_HI_DSpec}

\paragraph{Spectral Line Signal Transversally Colocated with the Foreground}\label{sec:1ps_FG_cosine_HI_colocated_DSpec}

As discussed in \S\ref{sec:1ps_FG_cosine_HI_colocated_spectrum}, despite the cosmic spectral line signal having a cosine-shaped spectral structure, its signatures are completely absent in the bispectrum phase angle. Therefore, the delay spectrum of the bispectrum phase considered in example~(\romannumeral 1)(a) is not expected to show any signatures in the Fourier domain as well.

\begin{figure*}
  \centering
  \subfloat[][Example~(\romannumeral 1)(b)\label{fig:EQ50_standard_bsp_dspec_1ps_FG_spindex_HI_displaced_achrAiry}]{\includegraphics[width=0.45\linewidth]{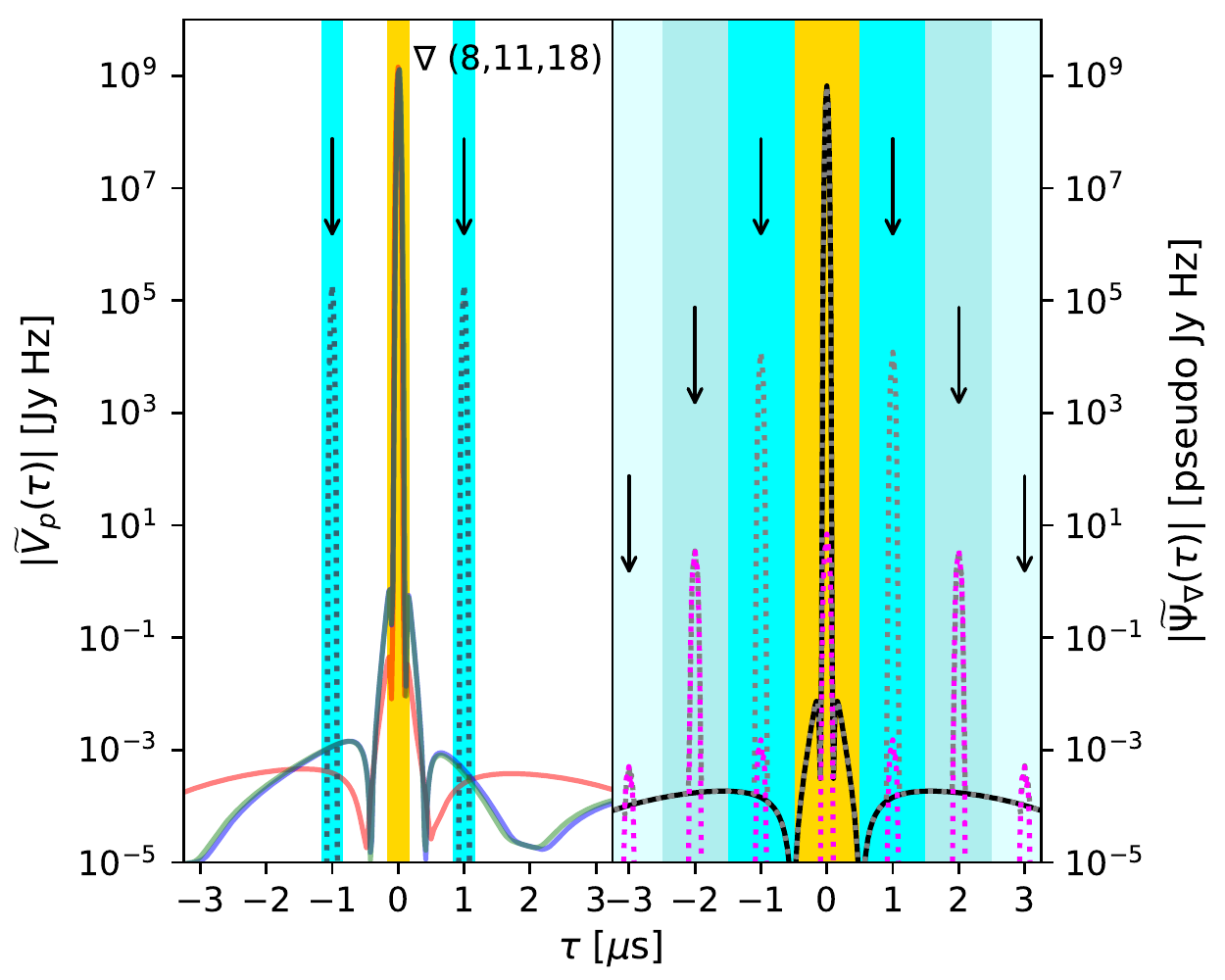}}
  \subfloat[][Example~(\romannumeral 2)\label{fig:EQ50_standard_bsp_dspec_GLEAM_FG_1ps_HI_achrAiry}]{\includegraphics[width=0.45\linewidth]{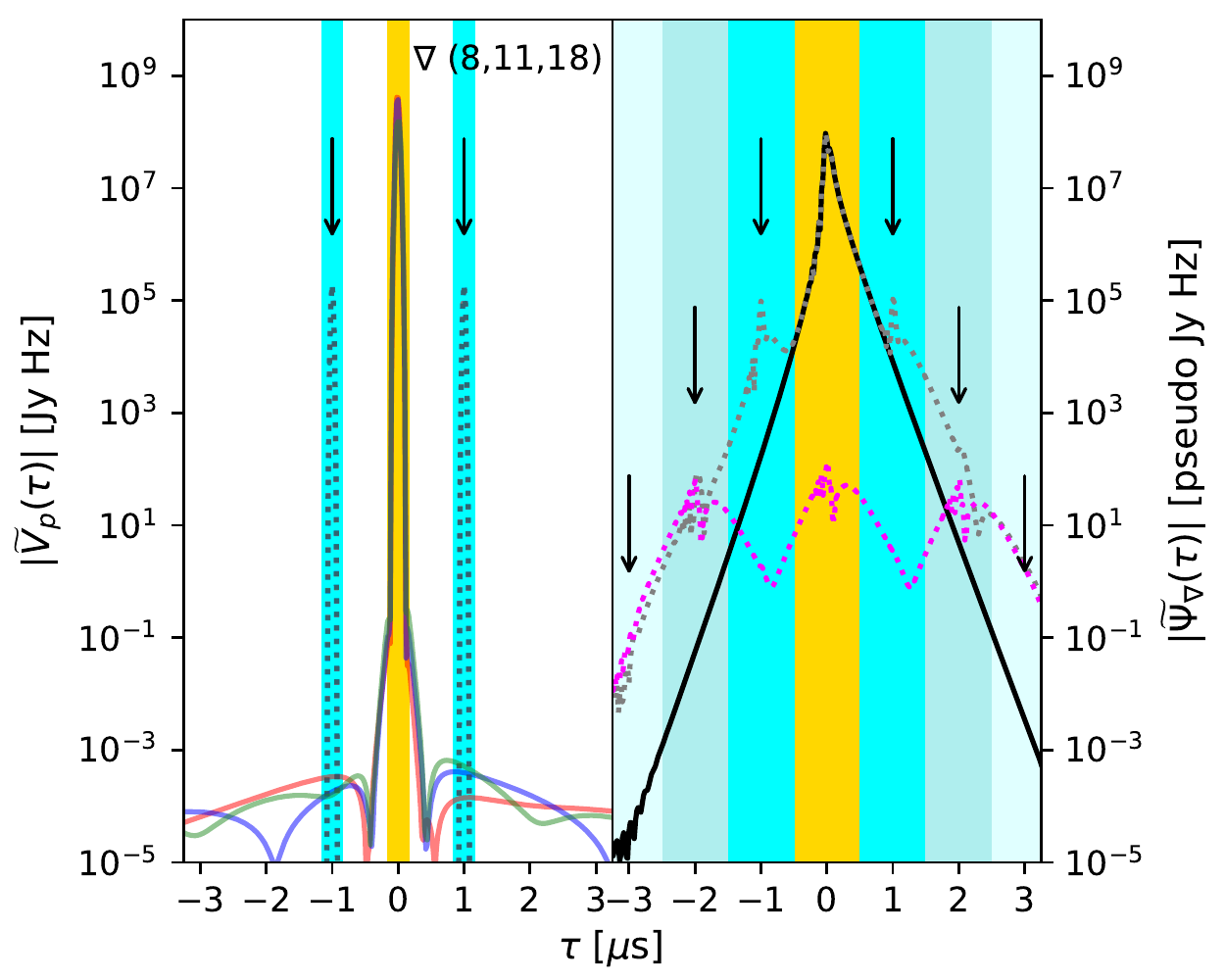}} \\
  \subfloat[][Example~(\romannumeral 3)\label{fig:EQ50_standard_bsp_dspec_1ps_FG_spindex_HI_21cmfast_achrAiry}]{\includegraphics[width=0.45\linewidth]{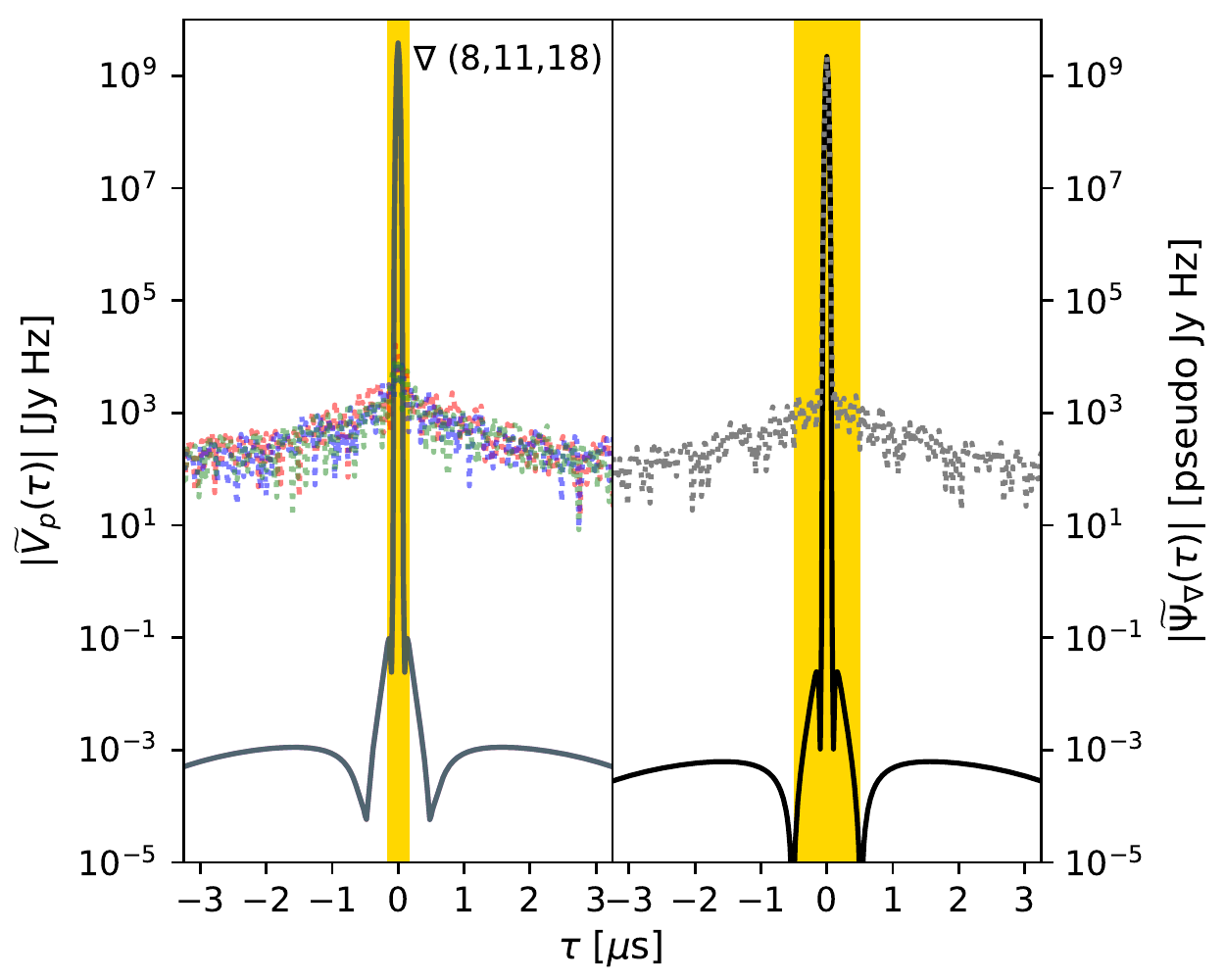}}
  \subfloat[][Example~(\romannumeral 4)\label{fig:EQ50_standard_bsp_dspec_GLEAM_FG_HI_21cmfast_achrAiry}]{\includegraphics[width=0.45\linewidth]{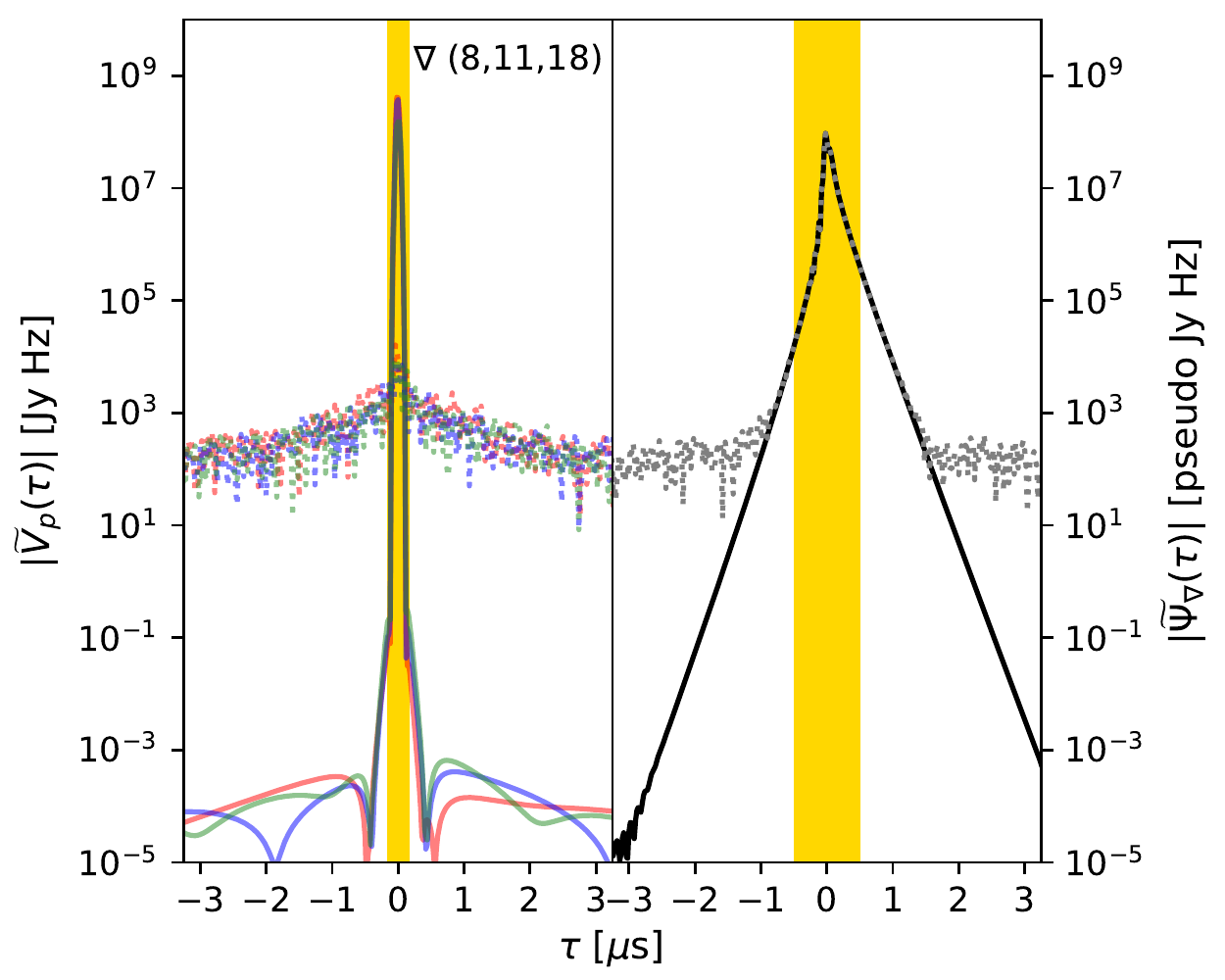}}
  \caption{Standard delay spectrum of the three visibilities, $\widetilde{V}_p(\tau)$ (red-, blue-, and green-colored curves in the left subpanels) and the bispectrum phase delay spectrum, $\widetilde{\Psi}_\nabla(\tau)$ (right subpanel) for the specified examples. The foreground and the cosmic \HI\ signatures in both subpanels are shown in solid and dotted curves, respectively. The 50.6~m equilateral triad used in these panels is specified at the top of the left subpanels. The yellow-shaded region denotes the foreground-dominated modes (\textit{foreground wedge}) in the delay spectra of the visibilities (left subpanels) and the bispectrum phase (right subpanels). The latter is wider due to triple convolution resulting from the multiplicative combination of the three visibility phases. In examples~(\romannumeral 1)(b) and (\romannumeral 2) that contain a point source cosmic \HI\ signal with a cosine-shaped spectrum, the downward arrows at $\tau=\pm\tau_\textrm{L}$ where the principal harmonic mode of the cosmic \HI\ signal is expected. The other downward arrows at $\tau=\pm 1\,\mu$s, $\tau=\pm 2\,\mu$s, $\tau=\pm 3\,\mu$s in the right subpanels denote the higher-order harmonics. Even higher-order harmonics are expected to be present but at negligible levels and are not shown. The pink dotted curves in examples~(\romannumeral 1)(b) and (\romannumeral 2) denote the difference between the delay spectra of the actual and the linear-order approximation of the bispectrum phase indicating that practically all the power in the second and third harmonics at $\tau=\pm 2\,\mu$s and $\tau=\pm 3\,\mu$s is entirely absent in the linear-order approximation. The second- and third-order terms also contribute at $\tau=0\,\mu$s and $\tau=\pm 1\,\mu$s respectively but are much smaller (by 10 and 7 orders of magnitude respectively) than that from the linear-order approximation at these harmonics. The cyan region denotes the possible range of offsets (same width as the \textit{foreground wedge}) that the expected delay mode the \HI\ signal could be subject to, i.e., $(\tau_\textrm{L}-\tau_{p,\textrm{h}}) \le |\tau_{p,\textrm{L}}^\textrm{obs}| \le (\tau_\textrm{L}+\tau_{p,\textrm{h}})$ instead of being precisely centered at $\tau_{p,\textrm{L}}^\textrm{obs}=\pm\tau_\textrm{L}$. The detailed analysis of these results for each of the examples is presented in the corresponding section in the text. In general, the foreground-dominated modes are wider in the bispectrum phase relative to the visibility delay spectra. Nevertheless, the cosmic spectral line signal is still detectable with a similar dynamic range and shape in the higher-order Fourier modes of the bispectrum phase compared to that in standard visibilities. A color version of this figure is available in the online journal. \label{fig:EQ50_vis_bsp_dspec_examples}}
\end{figure*}

The delay spectra of the rest of the sky model examples~(\romannumeral 1)(b), (\romannumeral 2), (\romannumeral 3), and (\romannumeral 4) are discussed below in detail and illustrated in Figure~\ref{fig:EQ50_vis_bsp_dspec_examples}.

\paragraph{Spectral Line Signal Transversally Displaced from Foreground}\label{sec:1ps_FG_cosine_HI_displaced_DSpec}

Figure~\ref{fig:EQ50_standard_bsp_dspec_1ps_FG_spindex_HI_displaced_achrAiry} shows the delay spectra of the visibilities (left subpanel) and the bispectrum phase (right subpanel) for example~(\romannumeral 1)(b). The region shaded in yellow in the left subpanel denotes the modes expected to be contaminated by the foregrounds, namely, the \textit{foreground wedge} \citep{bow09,liu09,liu14a,liu14b,dat10,liu11,gho12,mor12,par12b,tro12,ved12,dil13,pob13,thy13,dil14,thy15a,thy15b,thy16}, whose boundaries are determined by the horizon delay limit, $\tau_{p,\textrm{h}}=|\boldsymbol{b}_p|/c$. Since we use equilateral triads in these examples, $\tau_{p,\textrm{h}}=|\boldsymbol{b}_p|/c$ is identical for all $p$. The foreground delay spectra (solid lines) are seen to be the result of a convolution of a delta function at $\tau=0$ with the spectral window function's delay-domain response, $\widetilde{W}(\tau)$. The location of the peak of the foreground visibility delay spectra are displaced from $\tau=0\,\mu$s by a small amount $\delta\tau_p = \boldsymbol{b}_p\cdot\hat{\boldsymbol{s}}_\textrm{F}/c \lesssim 0.013\,\mu$s. Since this displacement is smaller than the resolution or the width of the response of the window function ($\delta\tau_p < \delta\tau_\textrm{w}$), the displacement is not discernible. The \HI\ spectrum which is cosine-shaped in this hypothetical example manifests as two delta functions at $\tau=\pm\tau_\textrm{L}=\pm 1\,\mu$s (expected at the locations of the downward arrows). Because the delay-domain response depends both on the transverse location and the frequency spectrum of the signal, the location of the delta functions of the cosine-shaped \HI\ spectrum in general could be subject to a delay offset that could be as large as the horizon delay limit, $\tau_{p,\textrm{h}}$. Thus, in general, depending on the location of the \HI\ signal, the delta functions corresponding to the cosmic \HI\ signal could be located anywhere in the cyan-shaded regions with $(\tau_\textrm{L}-\tau_{p,\textrm{h}}) \le |\tau_{p,\textrm{L}}^\textrm{obs}| \le (\tau_\textrm{L}+\tau_{p,\textrm{h}})$ instead of being precisely centered at $\tau_{p,\textrm{L}}^\textrm{obs}=\pm\tau_\textrm{L}$.

The foreground component of the delay spectrum of the bispectrum phase, $\widetilde{\Psi}_\nabla(\tau)$, in the right subpanel of Figure~\ref{fig:EQ50_standard_bsp_dspec_1ps_FG_spindex_HI_displaced_achrAiry} is very similar in magnitude, shape, and dynamic range to its counterpart in the standard delay spectrum, $\widetilde{V}_p(\tau)$ in the left subpanel. The general expectation for the foreground-contaminated modes is that they will be wider because the term, $\widetilde{\mathcal{E}}_\nabla^\textrm{F}(\tau)$, is derived from the convolution of the delay-transforms of the three visibility phase terms that appear as a product in the frequency spectrum. This triple product of visibility phase terms will, in general, widen the analogous \textit{foreground wedge} through the aforementioned convolution. Hence, the \textit{foreground wedge} in the delay spectrum of the bispectrum phase is determined by the sum of the three horizon limits in delay (Fourier) domain, $|\tau_{\nabla,\textrm{h}}|\le\sum_p \tau_{p,\textrm{h}}$, and is shown by the yellow-shaded region. The three shades of cyan denote the expected delay-offset locations of the \HI\ signal from the first- and higher-order harmonics of the cosine-shaped spectrum. The bright cyan shade denotes the range of delay offset around the expected location of the first harmonic centered around $(n\tau_\textrm{L}-\sum_p\tau_{p,\textrm{h}}) \le |\tau_{\nabla,\textrm{L}}^\textrm{obs}| \le (n\tau_\textrm{L}+\sum_p\tau_{p,\textrm{h}})$ with $n=1$ and primarily arises from the linear-order terms discussed in \S\ref{sec:bsp-linear-perturbations}.

The actual bispectrum phase will contain higher-order perturbations $\sim cos^n(2\pi f\tau_\textrm{L})$, and will not be captured by the linear-order approximation. The medium and pale shades of cyan regions correspond to the second and third harmonics ($n=2, 3$ respectively) expected from the second- and third-order perturbations respectively. The second-order terms are expected to contribute to both the zeroth and second harmonics since $\cos^2\theta\sim 1+\cos 2\theta$ (yellow and medium-cyan regions respectively) while the third-order terms will contribute to the first and third harmonics since $\cos^3\theta\sim 3\cos\theta+\cos 3\theta$ (dark- and pale-cyan regions respectively). There will be harmonics of even higher orders which have not been shown here because of their rapidly diminishing strengths. The width of each of these regions is the same as that of the yellow-shaded region.

Unlike when the \HI\ and foreground models are colocated and symmetric relative to each other in the transverse plane, the presence of non-zero phase angle fluctuations manifests prominently in the delay spectrum of the bispectrum phase (right subpanel) as sharply peaked functions (gray dotted lines) at $\tau\simeq \pm\tau_\textrm{L}=\pm 1\,\mu$s with small displacements around these locations $\sim\pm\tau_p$, where $\tau_p \lesssim 0.013\,\mu$s. These displacements are also indiscernible because $\delta\tau_p < \delta\tau_\textrm{w}$ and hence the 12 delta functions predicted in Equation~(\ref{eqn:HI-FG-PS}) have blended into two sharply-peaked functions one each on either side of $\tau=0\,\mu$s. Besides the principal first harmonics at $\tau\simeq\pm\tau_\textrm{L}$, sharp peaks are also seen as indicated by the downward arrows at $\tau\simeq\pm n\tau_\textrm{L}$ with $n=1,2,\ldots$ and $n=1$ being the principal (or first) harmonic. The pink dotted curve shows the difference between the delay spectrum of the actual bispectrum phase and that obtained with the linear-order approximation. The second and third harmonic components that arise from second- and third-order terms in the expansion of the bispectrum phase are lower relative to the first harmonic by factors $\sim 10^{-4}$ and $\sim 10^{-7}$ respectively which agree well with the fractional inaccuracy of the first-order prediction illustrated earlier in \S\ref{sec:bsp-linear-perturbations}. The second- and third-order terms also contribute to the zeroth and first harmonics that are not entirely represented by the linear-order approximation as indicated by the pink dotted curves around $\tau\simeq0\,\mu$s and $\tau\simeq\pm\tau_\textrm{L}$ respectively but these are negligible contributions (fractionally $\simeq 10^{-8}$) compared to the linear-order or actual values of the bispectrum phase delay spectra. There will be even higher-order harmonics in the actual bispectrum phase angles missed by the linear-order expansion but these contributions are expected to be even more increasingly negligible and are not shown.

\subsubsection{Example~(\romannumeral 2): Realistic Foreground and Unresolved Spectral Line Signal}\label{sec:GLEAM_FG_cosine_HI_DSpec}

Relative to the standard visibility delay spectra, $\widetilde{V}_p(\tau)$ in Figure~\ref{fig:EQ50_standard_bsp_dspec_1ps_FG_spindex_HI_displaced_achrAiry}, the visibility delay spectra for the GLEAM foreground model (left subpanel in Figure~\ref{fig:EQ50_standard_bsp_dspec_GLEAM_FG_1ps_HI_achrAiry}) appear wider and fill the \textit{foreground wedge} (central yellow region) as expected. The delay spectrum of the foreground component in the bispectrum phase (right subpanel of Figure~\ref{fig:EQ50_standard_bsp_dspec_GLEAM_FG_1ps_HI_achrAiry}) is much wider filling the correspondingly wider yellow central region due to $\widetilde{\mathcal{E}}_\nabla^\textrm{F}(\tau)$ which is formed by the triple convolution of the delay spectra of the visibility phase terms arising from the foregrounds as predicted by Eqs.~(\ref{eqn:cpdspec-line}) and (\ref{eqn:cpdspec-full-expanded}). The cosmic \HI\ signatures (gray dotted curves) are centered on the expected delays $\tau=\tau_\textrm{L}$ but as in the previous case, there are fainter copies at higher delay harmonics. As seen earlier, the higher-order harmonics that are not fully represented in the linear-order approximation contribute negligibly (by few to many orders of magnitude) to the zeroth and first harmonics (pink dotted curves). 

The most notable observation is that the cosmic \HI\ signatures appear to have the shape of the foreground signatures indicating they resulted from a convolution of a delta function with the foreground terms $\widetilde{\mathcal{E}}_\nabla^\textrm{F}(\tau)$ and $\widetilde{\gamma}_p(\tau)$ as detailed in Eqs.~(\ref{eqn:cpdspec-line}) and (\ref{eqn:cpdspec-full-expanded}). The previous example also had these effects but the foreground spectral signatures were not as rich to be clearly visible as in the present example. 

\subsubsection{Example~(\romannumeral 3): Unresolved Foreground and Fiducial Cosmic Spectral Line Signal} \label{sec:1ps_FG_21cmfast_HI_DSpec}

The left subpanel of Figure~\ref{fig:EQ50_standard_bsp_dspec_1ps_FG_spindex_HI_21cmfast_achrAiry} shows the standard visibility delay spectrum for the baselines comprising the 50.6~m equilateral triad (left subpanel) of the point source foreground and the fiducial EoR \HI\ model from the 21cmFAST simulations. On the right subpanel, the delay spectrum of the corresponding bispectrum phase is shown -- the foregrounds in solid black, and the combined foregrounds and EoR \HI\ fluctuations in dotted gray. Because of the simple spectral structure of the foregrounds, the delay spectra from either approach look remarkably similar to each other in their overall characteristics.

\subsubsection{Example~(\romannumeral 4): Realistic Foreground and Fiducial Cosmic Spectral Line Signal}\label{sec:GLEAM_FG_21cmfast_HI_DSpec}

Figure~\ref{fig:EQ50_standard_bsp_dspec_GLEAM_FG_HI_21cmfast_achrAiry} shows the delay spectra of the visibilities and the bispectrum phase. The foreground (solid red, green, and blue curves) and the fluctuating \HI\ components in the standard delay spectrum (left subpanel) are found to occupy the \textit{foreground wedge} (yellow shaded region) and extend beyond into the \textit{EoR window} respectively. The delay spectrum of the bispectrum phase (right subpanel) shows the foreground component (solid black curve) significantly wider due to the presence of significant spectral modes in the $\widetilde{\mathcal{E}}_\nabla^\textrm{F}(\tau)$ term arising out of the GLEAM foregrounds. This is in significant contrast to example~(\romannumeral 3) (see Figure~\ref{fig:EQ50_standard_bsp_dspec_1ps_FG_spindex_HI_21cmfast_achrAiry}) because the visibilities from the GLEAM foregrounds intrinsically contain more spectral structure than a single point source. The extent of this widening decreases with decreasing antenna spacings in the triad as will be shown in \S\ref{sec:mode-mixing} and thus will be less severe for a 14.6~m equilateral triad. The \HI\ component (dotted gray) does separate from the foregrounds at a level and shape similar to that in the standard delay spectrum. However, because the foreground component is significantly wider, the number of modes in which the \HI\ is detectable in the bispectrum phase approach is reduced to $|\tau| \gtrsim 1\,\mu$s relative to $|\tau| \gtrsim 0.12\,\mu$s in the standard delay spectrum of the visibilities.

In summary, the foregrounds occupy a larger range of inner spectral modes and hence the range of detectable cosmic spectral line signal modes are reduced but it is still significantly detectable in the remainder of the higher-order spectral modes. An alternate approach using the bispectrum phase angle is briefly outlined in \S\ref{sec:bsp-angle-approach} that could potentially avoid the widening of foreground contamination into larger spectral modes significantly.  

\section{Mode-Mixing in Bispectrum Phase}\label{sec:mode-mixing}

Mode-mixing in the context of spectral line experiments in the presence of foregrounds refers to the dependence of the line-of-sight spatial modes on the transverse spatial modes \citep{bow09,liu09,liu14a,liu14b,dat10,liu11,gho12,mor12,par12b,tro12,ved12,dil13,pob13,thy13,dil14,thy15a,thy15b,thy16} and is now commonly referred to as the \textit{foreground wedge}. To examine this effect in our bispectrum phase approach, we consider the 14.6~m and 50.6~m equilateral triads. The former samples lower order transverse spatial modes relative to the latter. In both cases, the foreground model is drawn from the GLEAM catalog, and the EoR \HI\ model from the fiducial 21cmFAST simulation of the EoR. 

\begin{figure}
\centering
  \subfloat[][Delay spectra of a 14.6~m equilateral triad \label{fig:EQ14_vis_bsp_dspec_GLEAM_FG_HI_21cmfast_achrAiry}]{\includegraphics[width=\linewidth]{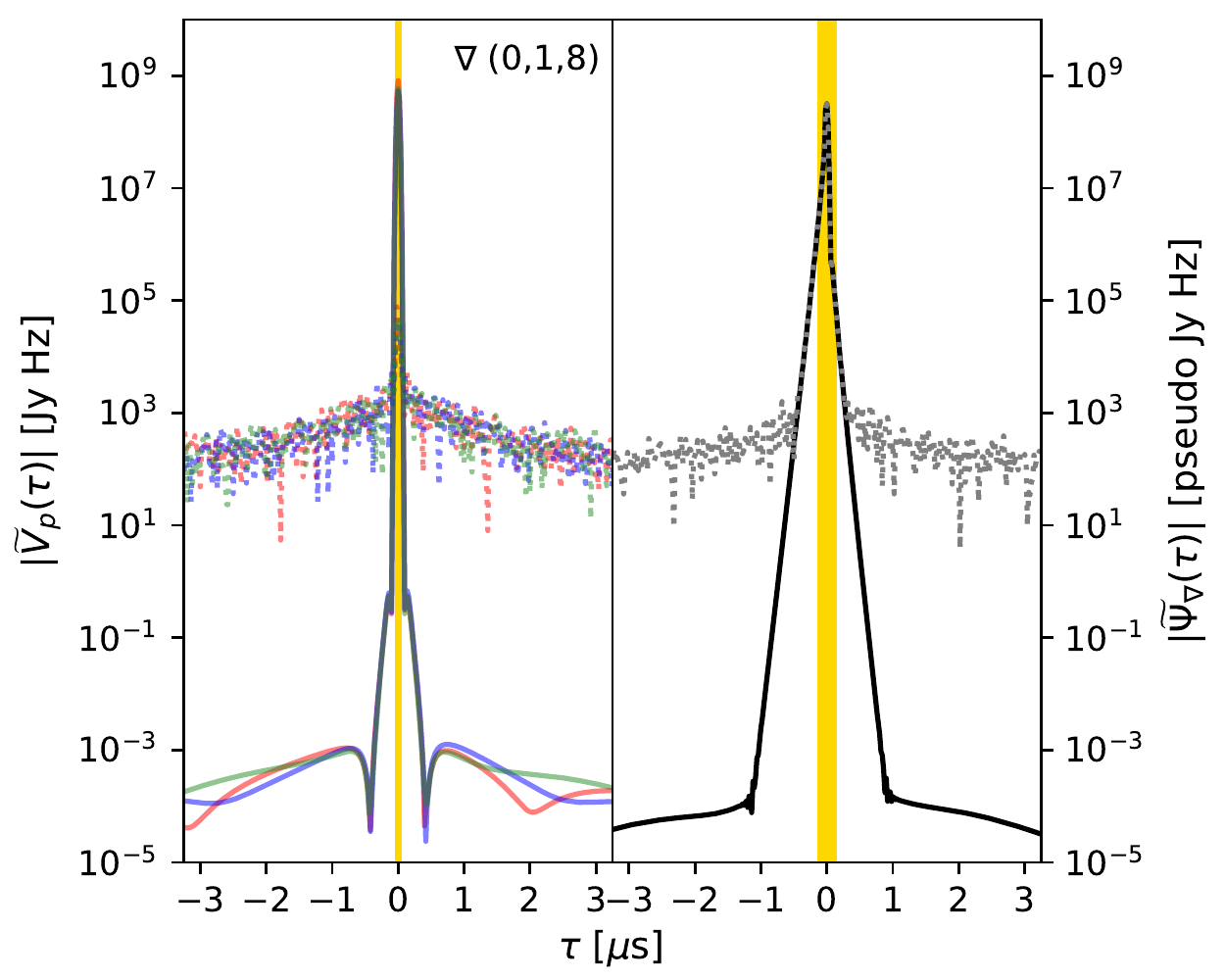}} \\
  \subfloat[][Delay spectra of a 50.6~m equilateral triad \label{fig:EQ50_vis_bsp_dspec_GLEAM_FG_HI_21cmfast_achrAiry_v2}]{\includegraphics[width=\linewidth]{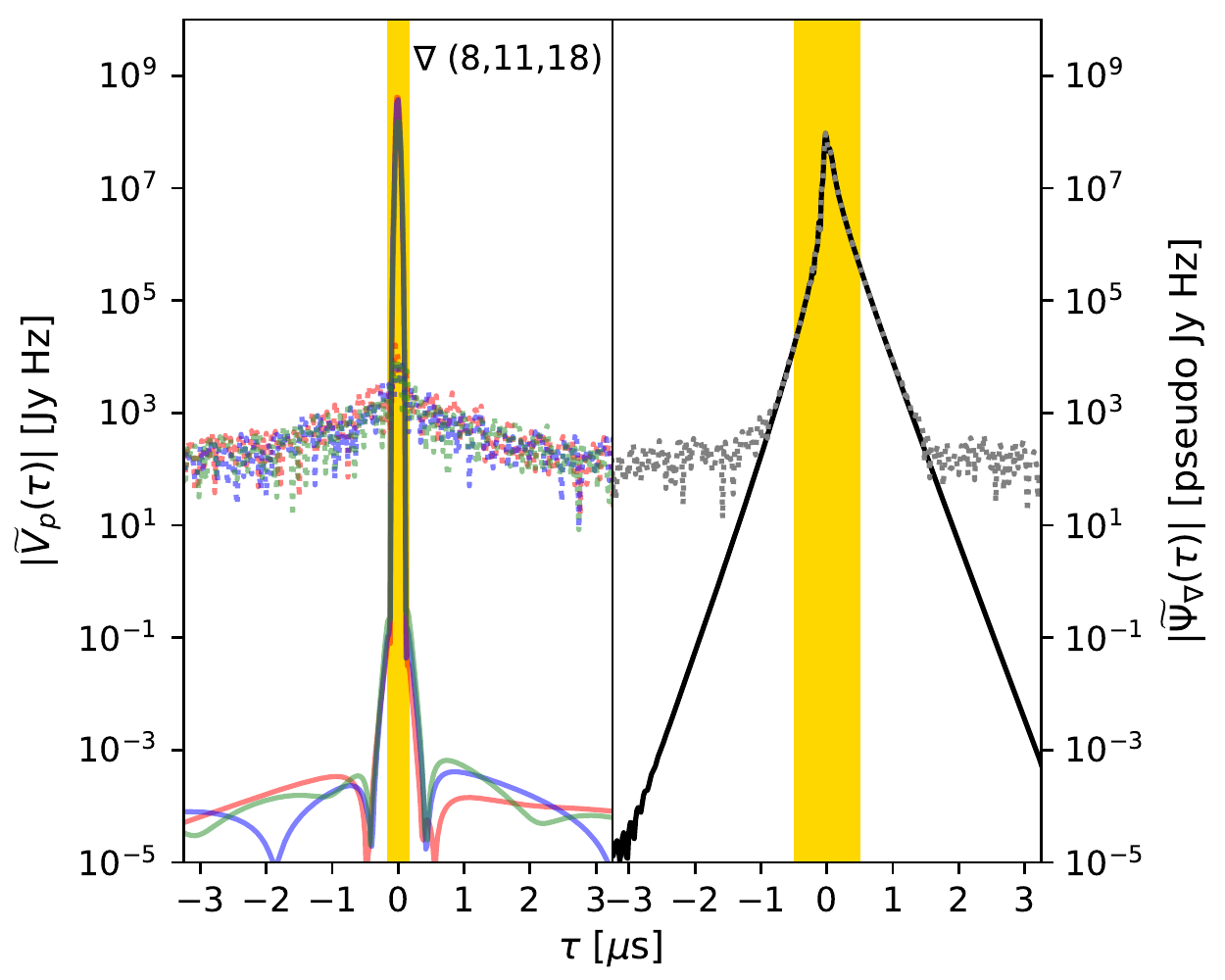}}
  \caption{\textit{Top:} Delay spectra of a 14.6~m equilateral triad. The left subpanel shows delay spectra of the three visibilities (red, green, and blue) comprising the triad for the foregrounds (solid curves) and the EoR \HI\ signal (dotted). The right subpanel shows the delay spectra of the bispectrum phase with foregrounds only (solid black curve), and with the EoR \HI\ fluctuations also present (gray dotted curve).  \textit{Bottom:} Same as Figure~\ref{fig:EQ14_vis_bsp_dspec_GLEAM_FG_HI_21cmfast_achrAiry} (top) but for a 50.6~m equilateral triad, and thus identical to Figure~\ref{fig:EQ50_standard_bsp_dspec_GLEAM_FG_HI_21cmfast_achrAiry}. Both the delay spectra of the visibilities (left subpanel) and the bispectrum phase (right subpanel) are wider in the case of the 50.6~m equilateral relative to the 14.6~m equilateral triad. This indicates that in the bispectrum phase, the transverse foreground modes also contaminate the line-of-sight foreground modes, as is the case in a standard delay spectrum approach using visibilities. Although cosmic signal-dominated modes are still accessible with a 50.6~m equilateral triad, they are much more accessible with a 14.6~m triad where the foreground contamination is much more tightly restricted. A color version of this figure is available in the online journal. \label{fig:mode_mixing_EQ14_EQ50_GLEAM_FG_HI_21cmfast_achrAiry}}
\end{figure}

Figure~\ref{fig:EQ14_vis_bsp_dspec_GLEAM_FG_HI_21cmfast_achrAiry} shows the delay spectra of the three (red, blue, green) visibilities (left subpanel) and of the bispectrum phase (right subpanel) corresponding to the 14.6~m equilateral triad. The solid curves correspond to the foreground component while the dotted curves represent the case when the cosmic \HI\ fluctuations are present. Figure~\ref{fig:EQ50_vis_bsp_dspec_GLEAM_FG_HI_21cmfast_achrAiry_v2} is the same but for a 50.6~m equilateral triad and is identical to  Figure~\ref{fig:EQ50_standard_bsp_dspec_GLEAM_FG_HI_21cmfast_achrAiry}. In both the delay spectra (left and right subpanels), the foreground component widens for the 50.6~m equilateral triad relative to the 14.6~m equilateral triad. This conclusively proves that the delay spectrum of the bispectrum phase is also subject to mode-mixing effects in general, wherein the transverse spatial modes contaminate the line-of-sight spatial modes, similar to the standard delay spectrum. However, the \textit{foreground wedge} is still limited in extent and the cosmic signal is detectable in the line-of-sight modes even for the 50.6~m equilateral triad. The foreground contamination is found to be much more limited and a wider range of cosmic signal-dominated modes are accessible with the usage of smaller triads such as the 14.6~m equilateral triad. \S\ref{sec:bsp-angle-approach} presents an outline of a variant to this approach using the bispectrum phase angle $\phi_\nabla^\textrm{m}(f)$ instead of $e^{i\phi_\nabla^\textrm{m}(f)}$ which is expected to be not as susceptible to mode-mixing as the latter is. 

\section{Impact of Foreground Spectral Characteristics}\label{sec:fg-impact}

Here, we compare the effects of the foreground spectral characteristics, such as the spectral index, on the delay spectra of the bispectrum phase and that of the visibilities. The example is similar to that in \S\ref{sec:1ps_FG_cosine_HI_displaced_spectrum}. It consists of a 100~Jy point-source foreground model $\approx 5$\arcdeg off-boresight and an unrealistic point-source \HI\ model at boresight of amplitude 10~mJy with a cosine-shaped spectrum of characteristic scale $\delta f_\textrm{L}=1/\tau_\textrm{L}=1$~MHz. The 50.6~m equilateral triad is used.

\begin{figure*}[htb]
\includegraphics[width=0.9\linewidth]{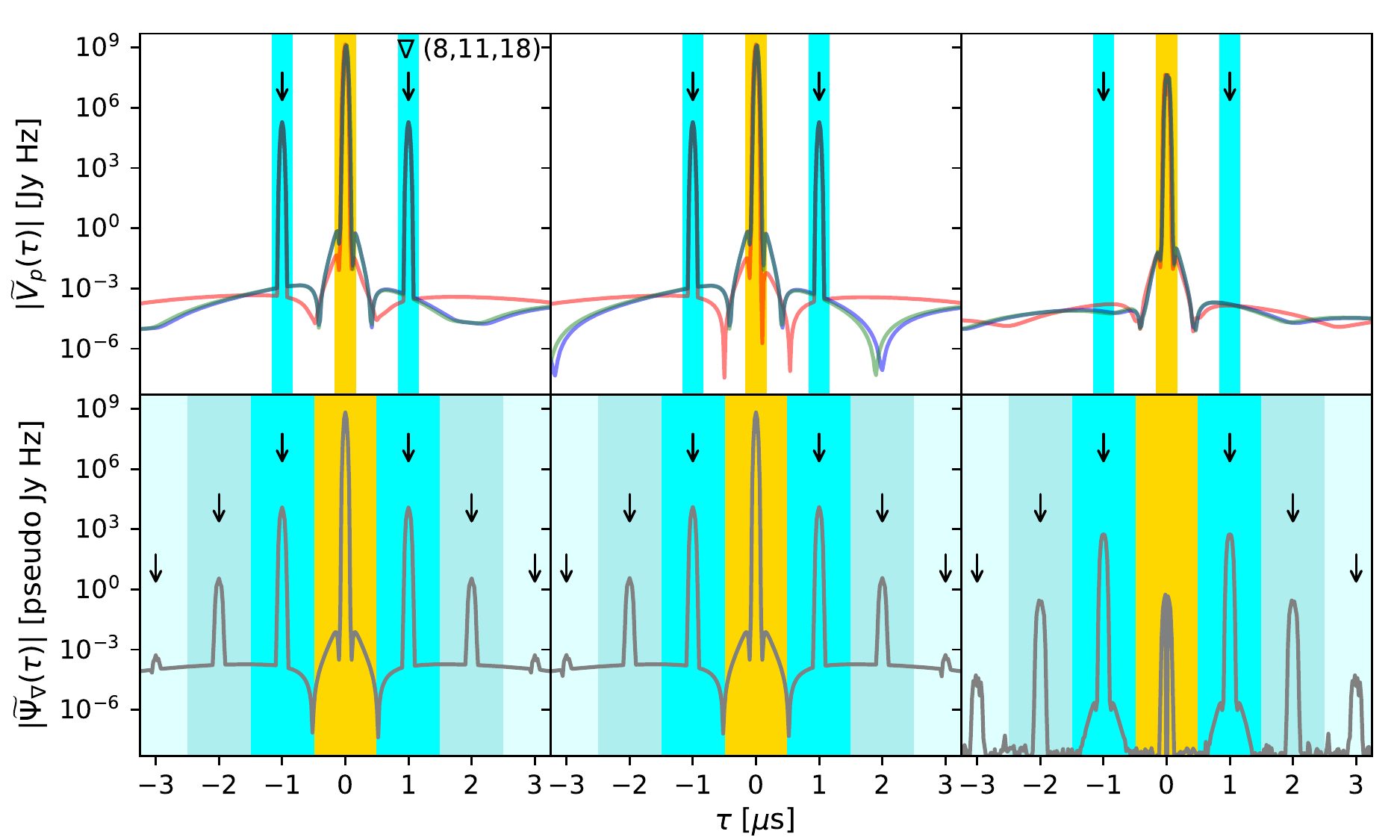}
\caption{Top panels correspond to the delay spectrum of the visibilities on a 50.6~m equilateral triad. Bottom panels correspond to the delay spectrum of the bispectrum phase on the same triad. The sky model used is similar to the hypothetical and unrealistic example~(\romannumeral 1)(b) with minor differences. The left and middle panels in both rows are derived using $\alpha=-0.8$ and $\alpha=0$ respectively for the point source foreground. The right panels shows the absolute value of the difference in the two delay spectra from these sky models. The downward arrows in the top panel indicate the expected location of the cosmic \HI\ signal at $\tau=\pm\tau_\textrm{L}$, whereas the downward arrows in the bottom panels show the location of the $n$-th harmonics $\tau=\pm\,n\tau_\textrm{L}$ as discussed in previous examples. The difference of the visibility delay spectra (top right subpanel) shows a residual that is reduced in magnitude but is purely foreground-based due to the differences in the spectral index of the two foreground models used and a complete absence of the \HI\ signatures at $\tau=\pm \tau_\textrm{L}$. The difference between the delay spectra of the bispectrum phase using the different spectral indices in the foreground models is shown in the bottom right subpanel. In contrast with the differenced visibility delay spectra (top-right subpanel), the \HI\ signatures do not vanish and the residuals have foreground-like signatures even at the harmonics where the \HI\ signatures are expected. This implies there is a mixing of the spectral characteristics of the foregrounds into the \HI\ signatures. A color version of this figure is available in the online journal. \label{fig:EQ50_effect_of_spindex_achrAiry}} 
\end{figure*}

The left panels of Figure~\ref{fig:EQ50_effect_of_spindex_achrAiry} use a spectral index, $\alpha=-0.8$, for the foreground model while the middle panels use $\alpha=0$. The right panels denote the absolute value of the difference between the delay spectra in the middle and the left panels. The top and bottom panels apply to the delay spectrum of the visibilities and the bispectrum phase respectively. In case of the latter, since different scalings to obtain the ``pseudo'' flux densities may have been applied, we normalized their peaks to be equal before the differencing. The downward arrows indicate where the \HI\ signatures are expected as sharply peaked functions. In the case of the bispectrum phase, the multiple downward arrows on each side of $\tau=0\,\mu$s denote the different harmonics of the $\tau_\textrm{L}=1/\delta f_\textrm{L}=1\,\mu$s spectral mode as discussed in the examples above. 

In the case of the standard visibility delay spectrum (top panels),  the \HI\ signatures are completely absent in the difference (right subpanel) and the peak of the foreground component has reduced by more than an order of magnitude. The difference still has a finite width around $\tau=0$ which can be attributed to the spectral index being different between the foreground models. This shows that the \HI\ signatures were not affected by the spectral characteristics of the foreground component and resulted in a perfect subtraction because of their additive behavior. The residual purely arises from the spectral characteristics of the two foreground models. 

In the case of the delay spectrum of the bispectrum phase (bottom panels), the differencing reduces the amplitude of the \HI\ signatures at the indicated harmonics but they do not vanish (right subpanel). Around each of these harmonics, the convolving effect of the foreground delay spectrum shapes can be seen. The fact that the \HI\ signatures at the harmonics do not vanish and the residuals at these harmonics contain the foreground-like signatures support the findings of Equation~(\ref{eqn:cpdspec-full-expanded}) wherein the spectral characteristics of the foregrounds are mixed with those of the cosmic \HI\ signal multiplicatively. 

\section{Summary}\label{sec:summary}

Numerous low-frequency radio interferometric measurements are underway to detect the large-scale distribution of baryons in the early Universe. This includes the detection of \HI\ using its 21~cm spectral line signature from high redshifts such as that from the \textit{Dark Ages}, the \textit{Cosmic Dawn}, the \textit{Epoch of Reionization}, and the \textit{Dark Energy}-dominated epoch in the Universe. These are expected to be extremely faint spectral signatures where the uncertainties are likely to be dominated by systematic uncertainties (especially of a spectral nature) from the instrument compounded by overwhelmingly bright and undesirable foreground emission from the Galaxy and extragalactic objects, rather than thermal noise in the detectors. One of the key challenges is the high-accuracy spectral calibration of the instrument which is typically required to have a fractional inaccuracy $\lesssim 10^{-5}$. The use of the bispectrum phase, which is independent of direction-independent antenna-based calibration and errors therein, has been presented as a viable alternative to statistically detect the presence of spectral line fluctuations. In this paper, we lay the foundational steps toward understanding the bispectrum phase in the context of the detection of faint cosmic spectral line fluctuations and examine its potential benefits and limitations.

The principal quantity investigated here for detection -- the bispectrum phase -- intrinsically measures the asymmetry (or dissimilarity) of the spatial distribution of the cosmic spectral line signal relative to the foregrounds in the transverse sky plane and is expressed as a rotation or fluctuation of the dominant phase angle from the bright foregrounds. In this paper, we focus on using the bispectrum phase to distinguish the faint cosmic spectral line fluctuations from the bright but spectrally smooth foregrounds along the spectral dimension, or equivalently along the line of sight. In the limit of small spectral fluctuations relative to the foregrounds, an approximate correspondence has been established between the approaches using the standard spatial coherence (visibilities) and that using the bispectrum phase. Specifically, the exact mathematical description for the spectral fluctuations in phase angles of both the visibilities and bispectrum have been established using a linear-order approximation (purely for analytical tractability) as being related to the ratio of the strength of the fluctuating signal to that of the foregrounds. Thus, existing Fourier domain techniques (e.g. \textit{delay transform}) can be readily employed to isolate these fluctuating spectral signatures of the cosmic signal. 

We have demonstrated that the linear-order approximation is a valid and useful handle to understand the bispectrum phase in the context of detecting faint spectral line fluctuations from cosmic structures. Although the linear-order approximation neglects the effects from the higher-order perturbation terms which manifest in actual measurements at higher harmonic modes in the delay spectrum, they are found to be negligible, typically by at least a few orders of magnitude. 

Except in virtually impossible scenarios where there is perfect symmetry of structure in the transverse sky plane between the source of spectral fluctuations and the foregrounds, the delay spectrum of the bispectrum phase corresponds well with that from the visibilities especially in aspects such as the foreground peak, the magnitude and shape of spectral line signatures, and the dynamic range between the two. This is confirmed using a wide variety of examples which ranged from simple point source models for foregrounds with zero or non-zero spectral index placed at boresight or off-boresight and a hypothetical and unrealistic point source \HI\ signal with a cosine-shaped spectrum to a realistic wide-area model of the foregrounds using the GLEAM catalog and a fiducial EoR model from 21cmFAST simulations. 

In the nearly impossible scenario that the transverse portion of the structures sourcing the spectral line fluctuations are perfectly symmetric relative to the foregrounds, the fluctuations in the bispectrum phase vanish and are undetectable even though there are clear spectral structures in the visibilities that will be detected in a standard delay spectrum. One of the key limitations of this approach stems from the fact that in the bispectrum phase, the foreground component contains a triple-product of the three interferometric visibility phase terms and this leads to a triply convolved and a wider response in the delay spectrum leading to higher levels of contamination in the low-order spectral modes thereby affecting detectability of the cosmic spectral line signal in these modes. In spite of this, the cosmic signal is still detectable on a wide range of Fourier modes of the bispectrum phase. A slightly modified approach using the bispectrum phase angles is also briefly presented in \S\ref{sec:bsp-angle-approach} that could potentially avoid this disadvantage to a significant extent. Further, since the bispectrum phase angle fluctuations depend on the foregrounds which are coupled multiplicatively, rather than additively, the spectral line signatures in a delay spectrum appear convolved with the delay response of the foreground spectral characteristics. Therefore, a straightforward interpretation of the spectral signatures seen in the delay spectrum is difficult and requires either a deconvolution approach to decouple the foreground effects or a detailed forward-modeling.  

Despite the limitations, the bispectrum phase approach is an intrinsic measure of the dissimilarity between the cosmic and the contaminating foreground structures and appears to be a viable, independent, and powerful tool to detect faint cosmic spectral line signatures in experiments where calibration of the instrument without corrupting the signatures of the cosmic signal is challenging. In a companion paper (Paper II), we present the first results from applying this technique to a small sample of data from the HERA telescope.

\begin{acknowledgments}
We acknowledge softwares including Python, Numpy, SciPy, Astropy, Matplotlib that made the numerical computations and the creation of figures presented in this manuscript possible. The National Radio Astronomy Observatory is a facility of the National Science Foundation operated under cooperative agreement by Associated Universities, Inc.
\end{acknowledgments}

\appendix

\section{An Alternate Approach with Bispectrum Phase Angles}\label{sec:bsp-angle-approach}

We briefly outline an alternate approach wherein the bispectrum phase angle, $\phi_\nabla^\textrm{m}(f)$, is used as the primary physical quantity of interest in deriving the results rather than using the complex Eulerian version, $e^{i\phi_\nabla^\textrm{m}(f)}$, as was done in the main text. We assume that the measured bispectrum phase angles are unwrapped accurately without introducing artefacts. In analogy to the complex Eulerian counterparts in \S\ref{sec:bsp-linear-perturbations}, we define:
\begin{align}
    V_\nabla(f) &= V_\textrm{eff}^\textrm{F}\,\phi_\nabla^\textrm{m}(f) \nonumber \\
    &= V_\textrm{eff}^\textrm{F}\,\left[\phi_\nabla^\textrm{F}(f) + \delta\phi_\nabla^\textrm{L}(f) + \delta\phi_\nabla^\textrm{N}(f)\right]  \label{eqn:visscale-angular} \\
    &= V_\textrm{eff}^\textrm{F}\, \Biggl[\phi_\nabla^\textrm{F}(f)+\frac{1}{2i}\sum_{p=1}^3\,\left(\frac{V_p^\textrm{L}(f)}{V_p^\textrm{F}(f)}-\frac{\conj{V_p^\textrm{L}}(f)}{\conj{V_p^\textrm{F}}(f)}\right) \nonumber \\ 
    &\qquad\qquad\qquad + \frac{1}{2i}\sum_{p=1}^3\,\left(\frac{V_p^\textrm{N}(f)}{V_p^\textrm{F}(f)}-\frac{\conj{V_p^\textrm{N}}(f)}{\conj{V_p^\textrm{F}}(f)}\right)\Biggr] \nonumber \\
    &= V_\nabla^\textrm{F}(f) + V_\nabla^\textrm{L}(f) + V_\nabla^\textrm{N}(f), \label{eqn:bsp-angle-equivalence}
\end{align}
where, 
\begin{align}
    V_\nabla^\textrm{F}(f) &= \phi_\nabla^\textrm{F}(f)\,V_\textrm{eff}^\textrm{F}, \\
    V_\nabla^\textrm{L}(f) &= \delta\phi_\nabla^\textrm{L}(f)\,V_\textrm{eff}^\textrm{F}, \\
    \textrm{and,}\quad V_\nabla^\textrm{N}(f) &= \delta\phi_\nabla^\textrm{N}(f)\,V_\textrm{eff}^\textrm{F}.
\end{align}
Combining the equations above, 
\begin{align}
    V_\nabla^\textrm{L}(f) &= \frac{1}{2i}\sum_{p=1}^3\,\left[\gamma_p^\textrm{F}(f)V_p^\textrm{L}(f)-\conj{\gamma_p^\textrm{F}}(f)\conj{V_p^\textrm{L}}(f)\right], \label{eqn:effective-line-visibilities-angular} \\ 
    V_\nabla^\textrm{N}(f) &= \frac{1}{2i}\sum_{p=1}^3\,\left[\gamma_p^\textrm{F}(f)V_p^\textrm{N}(f)-\conj{\gamma_p^\textrm{F}}(f)\conj{V_p^\textrm{N}}(f)\right]. \label{eqn:effective-noise-visibilities-angular}
\end{align}

Comparing Equation~(\ref{eqn:effective-line-visibilities-angular}) above with its analog, Equation~(\ref{eqn:effective-line-visibilities}), we notice that the key difference (disregarding some constants of proportionality) is the absence of the foreground bispectrum phase term, $e^{i\phi_\nabla^\textrm{F}(f)}$, in Equation~(\ref{eqn:effective-line-visibilities-angular}). Since this is the term that is a triple-product of the visibility phases in the triad that leads to a triple-convolution in the delay spectrum resulting in a significant widening of the \textit{foreground wedge} discussed in \S\ref{sec:examples} and \S\ref{sec:mode-mixing}, its absence in this alternate approach with bispectrum phase angles could potentially mitigate the contamination from \textit{mode-mixing} presented earlier to a substantial extent. This will be explored in detail in future work.

%

\end{document}